\documentclass[final,3p,times,doublespacing,fleqn]{elsarticle}
\biboptions{sort&compress}

\journal{Computer Methods in Applied Mechanics and Engineering}

\usepackage[active]{srcltx}
\usepackage{pdfsync}
\usepackage{hyperref}
\usepackage{multirow,setspace,booktabs}

\usepackage{amsthm,amssymb,array,float,latexsym,amsmath,amssymb,amsbsy,eurosym,theorem,amsfonts,mathrsfs,bbm,verbatim}

\usepackage[]{graphicx}

\usepackage{color}
\usepackage{showlabels}
\usepackage[utf8]{inputenc}

\renewcommand{\b}[1]{\boldsymbol{#1}} 

\newcommand{\lp}{\left(}
\newcommand{\rp}{\right)}

\newcommand{\lb}{\left[}
\newcommand{\rb}{\right]}

\newcommand{\p}{\partial}

\newcommand{\f}{\displaystyle\frac}

\renewcommand{\d}{\mbox{d}}

\newcommand{\grad}{{\nabla}}

\newcommand{\dyad}{\otimes}

\newcommand{\tmp}{{}^{\text{\scriptsize tmp}}} 

\newcommand{\ubar}{\overline{\b{u}}}
\newcommand{\upri}{\b{u}^\prime}
\newcommand{\intom}{\int_{V}\,I_{\Omega}\,}
\newcommand{\Abar}{\overline{A}}
\newcommand{\Apri}{A^\prime}

\newcommand{\cmark}{\checkmark}
\newcommand{\xmark}{\textbf{\sffamily X}}
\usepackage{tikz}
\usetikzlibrary{arrows.meta,positioning,shapes.geometric,shapes.misc}

\begin{document}

\begin{frontmatter}

\title{Reconstruction of glymphatic transport fields from subject-specific imaging data, with particular emphasis on cerebrospinal fluid flow and tracer conservation}

\author[oden]{A.~Derya~Bakiler}
\author[oden]{Michael J. Johnson}
\author[eind]{Michael R.A. Abdelmalik}
\author[oden]{Frimpong A. Baidoo} 
\author[baylor]{Andrew Badachhape}
\author[baylor]{Ananth V. Annapragada}
\author[oden]{Thomas J.R. Hughes}
\author[oden]{Shaolie S. Hossain\corref{cor}}

\address[oden]{Oden Institute for Computational Engineering and Sciences, The University of Texas at Austin, Austin, TX 78712, USA}
\address[eind]{Department of Mechanical Engineering, Eindhoven University of Technology, Groene Loper, 5612 AE Eindhoven, The Netherlands}
\address[baylor]{Department of Radiology,
Baylor College of Medicine, 1102 Bates Ave, Suite 850, Houston, TX 77030, USA}
\cortext[cor]{Corresponding author.}
\ead{shossain@oden.utexas.edu}

\begin{abstract} 
The reconstruction of physically valid transport fields from subject-specific imaging data is a fundamental challenge in image-based computational modeling due to measurement noise, modeling uncertainties, and discretization errors. 
Without a methodology to construct models that faithfully reflect the underlying physics, the mechanistic understanding of complex biological systems is inherently limited. 
In this work, we address this challenge within the context of the glymphatic system—the brain’s waste-clearance network—where cerebrospinal fluid (CSF) is transported through perivascular spaces and into the brain parenchyma to facilitate the removal of metabolic waste. 
We introduce a computational framework for the high-fidelity reconstruction of subject-specific glymphatic transport fields from spatiotemporal imaging data. 
The formulation utilizes an advection–diffusion model modified by a velocity decomposition that imposes mass conservation, enabling the recovery of solenoidal (divergence-free) velocity fields from noisy imaging data through the solution of a constrained inverse problem. 
The system is discretized using immersed isogeometric analysis with quadratic B-spline basis functions, which provides inherent regularization of imaging noise and ensures the continuity of derivatives of the inferred fields. 
We demonstrate the utility of the framework by obtaining spatially varying estimates of CSF velocity, diffusivity, and clearance parameters from contrast-enhanced magnetic resonance imaging data of tracer transport in a mouse brain. 
Forward simulations using the recovered fields show close agreement with experimental observations, validating the framework's ability to characterize complex transport dynamics while preserving physical integrity. 
This approach provides a generalizable methodology for the robust inference of physically consistent transport fields from imperfect imaging data, with broad applicability to the image-guided modeling of biological and engineering systems.

\end{abstract}


\end{frontmatter}

\tableofcontents
\section{Introduction}

Computational modeling has become an indispensable tool for investigating transport processes in complex biological systems, enabling systematic interrogation of governing mechanisms and parameters that are otherwise difficult to isolate or quantify \emph{in vivo}~\cite{hossain2012mathematical,hossain2013silico}. 
Traditional transport models, however, have largely been constrained by assumptions of spatial homogeneity and idealized geometry, compromising their physiological realism \cite{bohr_glymphatic_2022}. 
Historically, these simplifications were primarily driven by the prohibitive computational cost of resolving multiscale anatomical complexity and the scarcity of  high-resolution subject-specific volumetric imaging. 
Recent advances in medical imaging, coupled with improvements in  computational methods and power, have facilitated the shift toward image-based modeling frameworks that integrate subject-specific information directly into the numerical formulations, enabling the study of transport processes within physiologically realistic domains. 
However, the predictive capability of these models is contingent upon the reliable inverse inference of unknown transport processes and model parameters, such as  velocity, diffusivity and clearance rates, from sparse and noisy experimental imaging observations.

\begin{figure}[h!]
    \centering
    \includegraphics[width=0.9\textwidth]{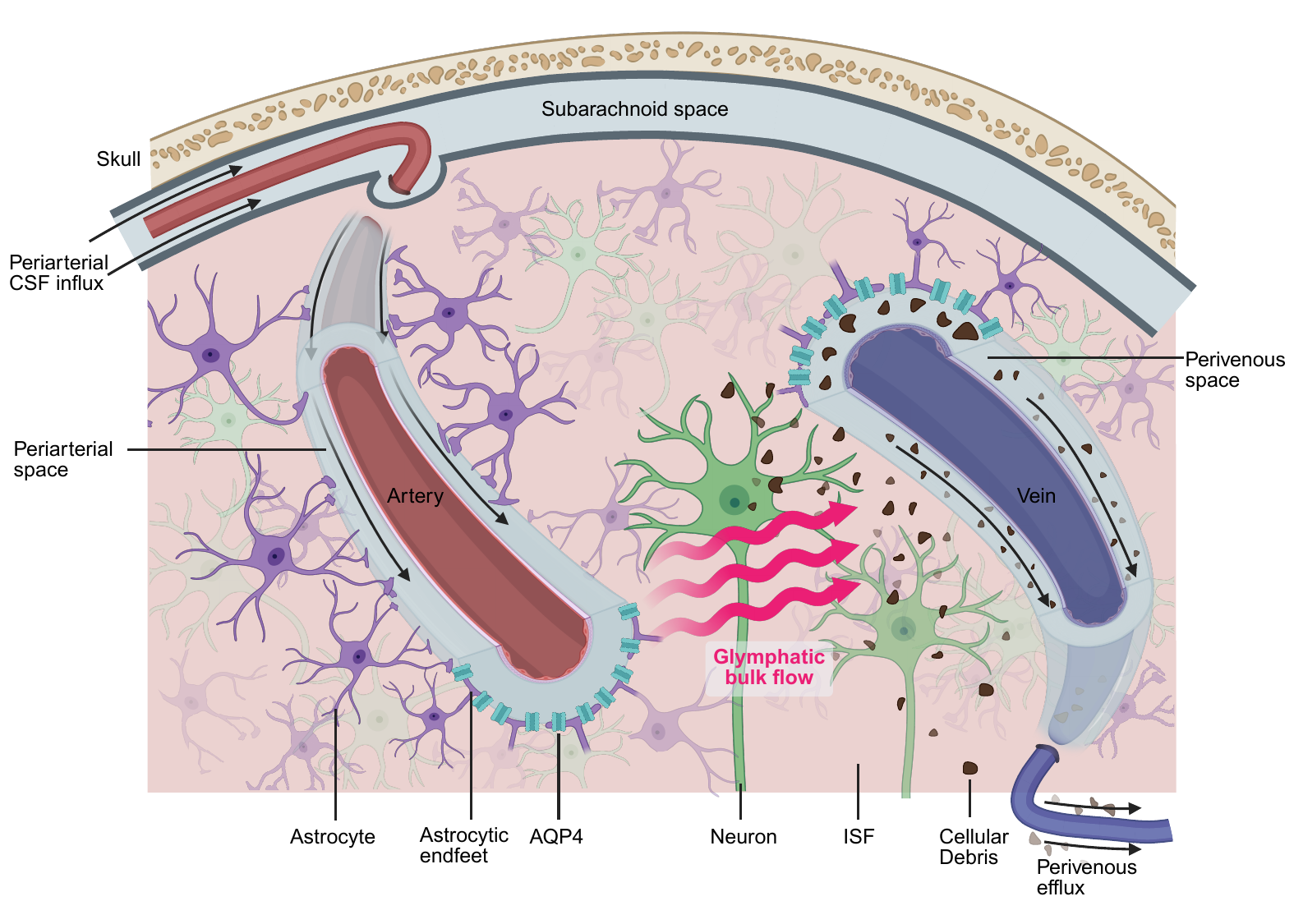}
    \caption{Schematic representation of glymphatic transport and clearance. Cerebrospinal fluid (CSF), produced by the choroid plexus via bulk transport from blood,  circulates through the subarachnoid space and enters the periarterial spaces of penetrating cortical arteries. CSF enters the brain parenchyma via transport mediated largely by aquaporin-4 (AQP4) water channels densely expressed on astrocytic endfeet. This influx facilitates the exchange of CSF with interstitial fluid (ISF), generating a bulk flow that traverses the interstitium. The resulting convective current flushes metabolic byproducts—including amyloid-beta and tau—toward perivascular spaces surrounding deep draining veins. The waste-laden fluid then exits the brain through lymphatic channels for systemic clearance.  This work focuses on modeling the transport dynamics within this coupled CSF–ISF system.
    Created with BioRender~\cite{biorender-glymph}.}
    \label{fig:glymph}
\end{figure}

The recently identified waste-clearance network of the brain, known as the glymphatic system~\cite{iliff_paravascular_2012}, presents a particularly challenging front for modeling transport, as the fundamental mechanisms governing glymphatic transport remain a subject of active debate ~\cite{mestre_brains_2020,guo_advection_2025}. 
This network of perivascular spaces facilitates the exchange of cerebrospinal fluid (CSF) and interstitial fluid (ISF) to eliminate metabolic byproducts, most notably amyloid-beta peptides~\cite{hablitz_glymphatic_2021}. 
Mechanistically, this occurs via multiscale transport involving  fluid movement along the narrow channels surrounding cerebral blood vessels  and through the brain tissue, as shown in Fig.~\ref{fig:glymph}. 
Delayed or impaired clearance of these waste proteins promotes their self-assembly into insoluble fibrils and eventual formation of large amyloid plaques—a pathological signature widely thought to initiate the neurodegenerative cascade leading to Alzheimer’s disease (AD) and other neurodegenerative diseases~\cite{macaulay_molecular_2021,buccellato_role_2022,hladky_glymphatic_2022}. 
While recent advances have enabled visualization of ventricular CSF flow dynamics~\cite{vikner_csf_2024},  glymphatic transport~\cite{iliff_brain-wide_2013,lee_brain_2020,wu_cerebrospinal_2025}, and neurodegenerative biomarkers~\cite{badachhape_pre-clinical_2020}, a comprehensive, brain-wide quantification of the transport and clearance of these solutes remains elusive.

Existing computational models are constrained by several key factors, including the simplification of complex anatomy into idealized 1D or axisymmetric geometries~\cite{asgari_glymphatic_2016}.  
In addition, the prohibitive cost of resolving intricate brain structures often necessitates reduced-order or homogenized approximations~\cite{solheim_geometry_2025}, which attenuate local velocity peaks and create scale mismatch between the discrete microscopic pathways of fluid transport and the coarse, voxel-averaged observations captured by clinical imaging~\cite{holter_interstitial_2017}. 
Furthermore, many models rely on the manual fitting of lumped transport parameters, such as effective tortuosity and pressure gradients, rather than direct inference of these fields from observed tracer dynamics~\cite{ray_analysis_2019}, thereby obscuring local transport heterogeneity and limiting predictive accuracy. 
While recent developments in regularized optimal mass transport (rOMT) have enabled 3D visualization of subject-specific glymphatic trajectories, these models are constrained by two major limitations: the assumption of spatially uniform, empirically-tuned diffusion, and the generation of velocity fields expressed in arbitrary units that lack the dimensional scaling necessary for physiological quantification~\cite{koundal_optimal_2020,benveniste_glymphatic_2021}.

Moreover, a consistent, mechanistic description of how waste is effectively taken out from the brain has yet to be established. 
Even though models that capture tracer dynamics within the brain exist, the parameters used to represent clearance are often only indirectly related to the underlying efflux mechanisms, limiting their physical interpretability~\cite{vinje_human_2023,chen_unbalanced_2024}.
Establishing a rigorous, anatomically faithful framework for quantitative characterization of glymphatic transport is therefore of significant interest~\cite{ding_impaired_2021,kopec_therapeutic_2025}; yet a persistent modeling gap remains: the reliable recovery of physically consistent, subject-specific transport fields from imaging data.
This challenge arises, in part, due to the disparate anatomical scales and the subtle pressure gradients involved, which necessitates high-fidelity numerical representations to resolve the resulting low-velocity regimes.

Despite the potential of image-based computational models to bridge this gap, ensuring physical consistency when deriving the unknown transport fields from raw imaging data is a nontrivial task. 
The calibration of these models to experimental observations is often complicated by measurement noise, tradeoffs between spatial and temporal resolution, and the need to sample and discretize continuous fields (e.g., flow)  from discrete, voxelized data. 
As a result, reconstructed velocity fields frequently reflect experimental artifacts rather than the underlying physics. 
Particularly consequential limitations in modeling glymphatic flow are lack of satisfaction of conservation of tracer concentration and incompressibility of the CSF. 
Conservation and incompressibility are thus basic hypotheses of our modeling.
Measurement and reconstruction errors in image-derived velocity fields commonly manifest as spurious divergence of the velocity field, leading to the violation of the conservation of mass. 
Therefore, enforcing divergence-free conditions is essential to maintain the integrity of numerical simulations.

These challenges frame the recovery of subject-specific transport fields as a constrained inverse problem: given noisy, discrete imaging observations, one must infer continuous, divergence-free velocity and transport fields that remain consistent with both the observed data and governing physical laws. 
Unlike purely data-driven or unconstrained reconstructions, such inverse formulations enable the systematic incorporation of prior physical knowledge to mitigate experimental artifacts and enhance physiological fidelity. 

Reconstruction of solenoidal (divergence-free) velocity fields from imaging data has been extensively explored in cardiovascular applications, particularly using phase-contrast magnetic resonance imaging (PC-MRI)~\cite{loecher_comparison_2012}. 
Established numerical approaches include: the projection of extracted velocity fields onto a divergence-free vector basis using finite difference schemes~\cite{song_noise_1993}; the integration of incompressibility constraints into post-processing routines via normalized convolution and divergence-free radial basis functions~\cite{busch_reconstruction_2013}; and the application of noise-reduction frameworks utilizing the divergence-free wavelet transform~\cite{ong_robust_2015}.
However, these methods developed for the high-momentum hemodynamic environment of cardiovascular flow are not directly applicable to the low-Péclet transport environment of the glymphatic system~\cite{asgari_glymphatic_2016}. 
In preclinical mouse brain models—the focus of the present work—glymphatic transport is typically captured via contrast-enhanced MRI (CE-MRI), requiring the inverse reconstruction of velocity fields from tracer evolution, rather than direct phase-velocity encoding available in PC-MRI. 
This process is complicated by two main factors: (i) CSF flow~\cite{mestre_flow_2018} is orders of magnitude slower than arterial blood flow ~\cite{meng_ultrafast_2022}, which drastically diminishes the signal-to-noise ratio and exacerbates existing spatiotemporal resolution limitations; and (ii) enforcing divergence-free constraints within image-based models is non-trivial, as many standard divergence-free discretizations rely on specialized finite element spaces and boundary-fitting mesh structures that may necessitate remeshing or geometric simplifications when applied to complex, image-derived domains. 
These challenges underscore the need for a robust, mass-conserving numerical framework for the inverse estimation of subject-specific transport fields and clearance parameters.

In this work, we present a computational framework for image-guided modeling of glymphatic transport that integrates divergence-free velocity reconstruction from imaging data with a 3D advection-diffusion formulation. We adapt the methodology of Hughes and Wells~\cite{hughes_conservation_2005}—originally developed to ensure conservation in advection-diffusion systems derived from Navier–Stokes simulations—to impose the incompressibility of the inferred CSF velocity fields. 
While the Hughes–Wells formulation is conventionally employed to maintain stability in forward solvers, we uniquely extend its utility to the inverse setting, where it serves as a physics-based regularizer to recover solenoidal fields from noisy image data. 
The resulting modified advection-diffusion equation is solved using an immersed finite element method within an isogeometric analysis (IGA) framework. 
This immersed approach enables the representation of complex brain morphology as an embedded object within a non-conforming mesh aligned to the image volume's bounding box, thereby bypassing the costly segmentation and mesh-generation steps required by traditional boundary-conforming methods. 
The glymphatic and CSF pathways are modeled via spatial heterogeneity in the image-inferred transport fields, such as velocity and diffusivity. 
Furthermore, IGA facilitates the projection of these high-dimensional image data and transport fields onto B-spline basis functions, naturally providing regularization and smoothing.
The smoothing properties of B-splines thus enable a reconstruction that preserves underlying biological variability, in contrast to voxel averaging which may end up homogenizing local dynamics.

The proposed framework establishes a physically grounded, subject-specific formulation that assimilates spatiotemporal CE-MRI observations to characterize the physical system while maintaining robustness to experimental noise and resolution limits. By synthesizing experimental observations with fundamental conservation laws, the present work achieves a degree of  mechanistic insight beyond what is possible through experimental parameterization alone. Although demonstrated here for glymphatic clearance, this framework is broadly applicable to a range of biological and engineering problems requiring the robust recovery of solenoidal advective fields from discrete experimental data.

The remainder of the manuscript is organized as follows.
Section~\ref{sec:gen} presents the governing equations for glymphatic transport, introducing a modified advection–diffusion formulation designed to ensure mass conservation even when the advective field is non-solenoidal. 
Section~\ref{sec:inverse} formulates the constrained inverse problem of recovering unknown transport parameters from image data. Following a discussion on the inherent limitations of the available experimental data, the modified formulation is extended to an inverse setting to enable subject-specific parameter estimation. 
Section~\ref{sec:resultsgen} presents numerical results for the full set of approaches developed in this work. Results are first shown for forward simulations using a baseline image-derived velocity field obtained through finite differencing, followed by results from the proposed inverse-inference framework for parameter recovery and validation.
Finally, Section~\ref{sec:conc} provides concluding remarks and outlines potential directions for future research.

\section{Forward advection-diffusion model with conservation}\label{sec:gen}

In this section, we first present the governing advection–diffusion equations as applied to glymphatic transport. 
This is followed by establishing the weak-forms for the conservative, advective, and \emph{modified advective} forms of the advection-diffusion equation.
We conclude this section by providing details on the numerical implementation of the modified advective forward model using immersed IGA and finite elements.  

\subsection{Governing equations and forward model definition}\label{sec:goveq}

Let $\Omega$ denote the volume of a mouse brain, with the boundary denoted as $\Gamma$.
The geometry of the brain volume is obtained by segmenting the brain parenchyma and meninges from contrast-enhanced magnetic resonance imaging (CE-MRI) tracer data. 
These tracer experiments involve intrathecal injection of a water-soluble contrast-agent into the CSF pathways of the mouse brain, followed by serial imaging  to monitor the clearance of the tracer through the glymphatic system.
Because endogenous soluble amyloid cannot be tracked\textit{ in vivo}, exogenous contrast agents are utilized as surrogates to characterize spatiotemporal dynamics of waste transport within the glymphatic system. 
Thus, let $c(x,y,z,t)$ denote the concentration of tracer in the CSF, in nM, where nM is nanomolar concentration. 
The glymphatic transport of the tracer is then modeled by the following equation
\begin{equation}\label{eq:strong}
    \begin{aligned}
        \f{\p c}{\p t} = \nabla \cdot (D \nabla c) - \nabla \cdot (c \b{u}) \quad \text{in} \ \Omega \,.
    \end{aligned}
\end{equation}
The partial differential equation (PDE) in~(\ref{eq:strong}) is an advection-diffusion equation, where $D(x,y,z)$ denotes the diffusivity of soluble tracer molecules in mm$^2$/min, and $\b{u}(x,y,z)$ denotes the CSF velocity field in mm/min. 
The clearing out of the tracer from the brain via the glymphatic drainage pathways is accounted for by the Robin boundary condition 
\begin{equation}\label{eq:BCs}
    \begin{aligned}
        - D \, \nabla c \cdot \b{n} = \gamma \, c \quad \text{on} \,\, \Gamma \,,
    \end{aligned}
\end{equation}
where $\b{n}$ is the outward unit normal vector on $\Gamma$, and $\gamma(x,y,z)$ is the clearance parameter in mm/min, which varies over $\Gamma$. 
The boundary condition~(\ref{eq:BCs}) assumes that the diffusive flux of clearing tracer is proportional to the tracer concentration at that point on the boundary, and that clearance only takes place along the boundary of the domain.
A greater value for the clearance parameter $\gamma$ indicates higher clearance, and vice versa. 
Finally, the initial condition reads
\begin{equation}\label{eq:ICs}
    \begin{aligned}
        c \, (x,y,z, 0) = \hat{c} \, (x,y,z) \quad \text{on} \,\, \Omega \,, \\
    \end{aligned}
\end{equation}
where $\hat{c}$ denotes the image tracer intensity at an initial time instant. 
The velocity field $\b{u}$ is constrained with the zero-velocity boundary condition, $\b{u} = \b{0}$ on $\Gamma$.
This is accompanied by the assumption that $\b{u}$ is divergence-free.
The spatial variation of the velocity $\b{u}$, diffusivity $D$, and clearance parameter $\gamma$ remain to be determined. 

\subsection{Classical and modified advection–diffusion weak forms}

We begin by defining a residual $R(c)$ by rewriting Eq.~(\ref{eq:strong}) as
\begin{equation}\label{eq:R}
    \begin{aligned}
       R(c) := \f{\p c}{\p t} - \nabla \cdot (D \nabla c) + \nabla \cdot (c \b{u}) = 0 \,.
    \end{aligned}
\end{equation}
The weak form of the equation in conservative form is then obtained by multiplying the residual form~(\ref{eq:R}) with a test function $w \in H^1(\Omega) $ and integrating over the domain $\Omega$. 
A weak form, referred to as the \emph{standard conservative form} with streamline upwind Petrov-Galerkin (SUPG)~\cite{brooks_streamline_1982} stabilization, is to find $c \in H^1 (\Omega)$ such that the following holds: $\forall w \in H^1(\Omega)$,
\begin{equation}\label{eq:weak-cons}
    \begin{aligned}
       \f{d}{dt} \int_{\Omega} w \, c \, \d x + \int_{\Omega} D \, \nabla \, w \cdot \nabla c \, \d x + \int_{\Omega} w \,\nabla \cdot  (c \b{u}) \, \d x  
       &+ \int_{\Gamma} w \, \gamma \, c \, \d\Gamma    + \int_{\Omega} \sigma \, R(c) \, \b{u} \cdot \nabla w \,\d x = 0 \,,
    \end{aligned}
\end{equation}
where $\sigma$ is the stabilization parameter defined as
\begin{equation}\label{eq:weak-supg-sigma}
    \begin{aligned}
        \sigma &:= \f{h}{2 |\b{u}|} \min\lp \f{\text{Pe}^{\,h}}{3}, \,1 \rp \,, \quad \text{Pe}^{\,h} := \f{|\b{u}|\,h}{2 D}\\
\end{aligned}
\end{equation}
where $h$ is a characteristic length for the element, and ${\text{Pe}}^{\,h}$ is the element Peclet number.
Whether the weak form in Eq.~(\ref{eq:weak-cons}) conserves the quantity $c$ can be checked by first integrating the advective term by parts, and invoking the zero-velocity boundary condition on $\b{u}$,
\begin{equation}\label{eq:weak-cons-ibp}
    \begin{aligned}
       \f{d}{dt} \int_{\Omega} w \, c \, \d x + \int_{\Omega} D \, \nabla \, w \cdot \nabla c \, \d x - \int_{\Omega} \nabla w \,\cdot  (c \b{u}) \, \d x  
       &+ \int_{\Gamma} w \, \gamma \, c \, \d\Gamma    + \int_{\Omega} \sigma \, R(c) \, \b{u} \cdot \nabla w \,\d x = 0 \,,
    \end{aligned}
\end{equation}
and then setting $w \equiv 1 $, since $1 \in H^1 (\Omega)$ as well,
\begin{equation}\label{eq:weak-cons-w1}
    \begin{aligned}
       \f{d}{dt} \int_{\Omega}  c \, \d x  + \int_{\Gamma} \gamma \, c \, \d\Gamma  = 0 \,.
    \end{aligned}
\end{equation}
Eq.~(\ref{eq:weak-cons-w1}) shows that any change in the total concentration over the domain in time $dt$ must be balanced by whatever is taken out from the boundaries, demonstrating conservation of $c$. 
However, the advective term in the weak form~(\ref{eq:weak-cons}), which contains the divergence of the velocity field $\b{u}$, proves to be problematic for non-solenoidal $\b{u}$. 
Regions of positive and negative divergence in $\b{u}$ act as spurious sinks and sources in the domain, leading to nonphysical accumulation in some locations and negative concentrations in others, which may result in unbounded growth or decay of the numerical solution.

Rewriting the equation in advective form represents a viable alternative formulation, offering improved numerical stability and better behavior under non-divergence-free flow~\cite{hughes_conservation_2005}.
The weak form in \emph{standard advective form} with SUPG stabilization is to find $c \in H^1 (\Omega)$ such that the following holds: $\forall w \in H^1(\Omega)$, 
\begin{equation}\label{eq:weak-advec}
    \begin{aligned}
       \f{d}{dt} \int_{\Omega} w \, c \, \d x + \int_{\Omega} D \, \nabla \, w \cdot \nabla c \, \d x + \int_{\Omega} w \, \b{u} \cdot \nabla c \, \d x 
       + \int_{\Gamma} w \, \gamma \, c \, \d x  + \int_{\Omega} \sigma \, R(c) \, \b{u} \cdot \nabla w \,\d x = 0 \,,
    \end{aligned}
\end{equation}
where the following identity has been used,
\begin{equation}\label{eq:divcu-identity}
    \begin{aligned}
        \nabla \cdot (c \b{u}) = (\nabla \cdot \b{u}) \, c + \b{u} \cdot \nabla c \,,
    \end{aligned}
\end{equation}
followed by the assumption that $\nabla \cdot \b{u} = 0 $.  

Note that for a solenoidal velocity field, this assumption is true and the conservative form~(\ref{eq:weak-cons}) and advective form~(\ref{eq:weak-advec}) are identical.
However, velocity fields obtained from image data rarely meet this assumption, leading to the loss of conservation. 
To demonstrate this, we set $w\equiv1$ in the standard advective form in Eq.~(\ref{eq:weak-advec}),
\begin{equation}\label{eq:weak-advec-w1}
    \begin{aligned}
       \f{d}{dt} \int_{\Omega} c \, \d x + \int_{\Omega} \b{u} \cdot \nabla c \, \d x 
       + \int_{\Gamma} \gamma \, c \, \d\Gamma = 0 \,,
    \end{aligned}
\end{equation}
showing that the concentration $c$ is not conserved unless the second term in Eq.~(\ref{eq:weak-cons-w1}) vanishes.

In order to reconcile the conservation advantage of~(\ref{eq:weak-cons}) and the stability advantage of~(\ref{eq:weak-advec}), we introduce a multiscale-inspired decomposition of the velocity as outlined by Hughes and Wells (HW) in~\cite{hughes_conservation_2005} to build a stable advection-diffusion scheme that conserves the quantity $c$ even when the velocity field is non-solenoidal.
The velocity field is decomposed into a \lq\lq coarse\rq\rq $\,$ scale component $\ubar$, and a \lq\lq fine\rq\rq$\,$ scale component $\upri$,
\begin{equation}\label{eq:advec-decomp}
    \begin{aligned}
        \b{u} := \ubar + \upri \quad \text{where} \quad \upri = 0 \, \text{on} \, \Gamma \,,
    \end{aligned}
\end{equation}        
where we have assumed that the fine scale component is zero on the boundaries, or equivalently that $\b{u} = \ubar$ on $\Gamma$.
The field $\ubar$ can also be interpreted as the measured or data-derived velocity, with $\upri$ representing a correction to $\ubar$ that ensures a divergence-free velocity field $\b{u}$.
Inserting the decomposition~(\ref{eq:advec-decomp}) into the advective weak form~(\ref{eq:weak-advec}) yields the \emph{Hughes-Wells (HW) modified advective weak form}, find $c \in H^1 (\Omega)$ such that the following holds $\forall w \in H^1(\Omega)$,
\begin{equation}\label{eq:weak-advec-HWC}
    \begin{aligned}
       \f{d}{dt} \int_{\Omega} w \, c \, \d x + \int_{\Omega} D \, \nabla \, w \cdot \nabla c \, \d x &+ \int_{\Omega} w \, \ubar \cdot \nabla c \, \d x + \int_{\Omega} w \, \upri \cdot \nabla c \, \d x 
       + \int_{\Gamma} w \, \gamma \, c \, \d\Gamma \\
       &+ \int_{\Omega} \sigma \, R(c) \, \ubar \cdot \nabla w \,\d x + \int_{\Omega} \sigma_p \, R_p(c) \,\upri \cdot \nabla w \,\d x = 0 \,,
    \end{aligned}
\end{equation}
where the residuals are defined as
\begin{equation}\label{eq:weak-advec-decomp-supg}
    \begin{aligned}
       R(c) &:= \f{\p c}{\p t} - \nabla \cdot (D \nabla c) + \ubar \cdot \nabla c \,, \\
       R_p(c) &:= \upri \cdot \nabla c \,,\\
    \end{aligned}
\end{equation}
and the stabilization parameters $\sigma$ and $\sigma_p$ are
\begin{equation}\label{eq:weak-advec-decomp-supg-sigma}
    \begin{aligned}
        \sigma &:= \f{h}{2 |\ubar|} \min\lp \f{\overline{\text{Pe}}^{\,h}}{3}, \,1 \rp \quad \text{where} \quad  \overline{\text{Pe}}^{\,h} := \f{|\ubar|\,h}{2 D} \,, \\
        \sigma_p &:= \f{h}{2 |\upri|} \min\lp \f{{\text{Pe}^\prime}^{h}}{3}, \,1 \rp \quad \text{where} \quad  {\text{Pe}^\prime}^{h} := \f{|\upri|\,h}{2 D} \,.
\end{aligned}
\end{equation} 
The term $R_p(c)$ is a residual in the following sense: if $\ubar$ was the exact divergence-free velocity field, $\upri$ would be zero, so $\upri$ itself may be considered a residual, hence so is $R_p(c)$.
Nevertheless, in practice, $\upri \neq 0$, and thus it needs stabilization. 
Note that this modification is consistent with Eq.~\ref{eq:strong}, in the sense that the solutions to Eq.~(\ref{eq:strong}) also satisfy Eq.~(\ref{eq:weak-advec-HWC}) when $\b{u}$ is solenoidal.
In a manner similar to the previously established standard forms, we check the HW modified advective weak form~(\ref{eq:weak-advec-HWC}) for conservation.
Recalling the identity given in Eq.~(\ref{eq:divcu-identity}), we can rewrite the second and third terms of Eq.~(\ref{eq:weak-advec-HWC}) such that they now read
\begin{equation}\label{eq:weak-advec-decomp-play}
    \begin{aligned}
       \f{d}{dt} \int_{\Omega} w \, c \, \d x &+ \int_{\Omega} D \, \nabla \, w \cdot \nabla c \, \d x + \int_{\Omega} w \, \nabla \cdot (c \ubar) \, \d x - \int_{\Omega} w \, (\nabla \cdot \ubar) \,c\, \d x + \int_{\Omega} w \, \upri \cdot \nabla c \, \d x \\
       &+ \int_{\Gamma} w \, \gamma \, c \, \d\Gamma 
       + \int_{\Omega} \sigma \, R(c) \, \ubar \cdot \nabla w \,\d x + \int_{\Omega} \sigma_p \, R_p(c) \,\upri \cdot \nabla w \,\d x = 0 \,.
    \end{aligned}
\end{equation}
Integrating the third term by parts yields 
\begin{equation}\label{eq:weak-advec-decomp-play-1}
    \begin{aligned}
       \f{d}{dt} \int_{\Omega} w \, c \, \d x &+ \int_{\Omega} D \, \nabla \, w \cdot \nabla c \, \d x 
       - \int_{\Omega} \nabla w \cdot (c \, \ubar) \, \d x + \int_{\Gamma} w \, c\, \ubar_n \, \d x - \int_{\Omega} w \, (\nabla \cdot \ubar) \,c\, \d x + \int_{\Omega} w \, \upri \cdot \nabla c \, \d x \\
       &+ \int_{\Gamma} w \, \gamma \, c \, \d\Gamma
       + \int_{\Omega} \sigma \, R(c) \, \ubar \cdot \nabla w \,\d x + \int_{\Omega} \sigma_p \, R_p(c) \,\upri \cdot \nabla w \,\d x = 0 \,,
    \end{aligned}
\end{equation}
where $\ubar_n := \ubar \cdot \b{n}$. 
The boundary term coming from integration by parts disappears since the zero-velocity boundary condition on $\b{u}$ implies $\b{u} = \b{\ubar} = 0$ on $\Gamma$, which renders the following alternative form of Eq.~(\ref{eq:weak-advec-HWC})
\begin{equation}\label{eq:weak-advec-decomp-play-2}
    \begin{aligned}
       \f{d}{dt} \int_{\Omega} w \, c \, \d x &+ \int_{\Omega} D \, \nabla \, w \cdot \nabla c \, \d x 
       - \int_{\Omega} \nabla w \cdot (c \, \ubar) \, \d x - \int_{\Omega} w \, (\nabla \cdot \ubar) \,c\, \d x + \int_{\Omega} w \, \upri \cdot \nabla c \, \d x \\
       &+ \int_{\Gamma} w \, \gamma \, c \, \d\Gamma 
       + \int_{\Omega} \sigma \, R(c) \, \ubar \cdot \nabla w \,\d x + \int_{\Omega} \sigma_p \, R_p(c) \,\upri \cdot \nabla w \,\d x = 0 \,.
    \end{aligned}
\end{equation}
Conservation of the quantity $c$ is then checked by setting $w\equiv1$ in Eq.~(\ref{eq:weak-advec-decomp-play-2}),
\begin{equation}\label{eq:weak-advec-decomp-play-2-w1}
    \begin{aligned}
       \f{d}{dt} \int_{\Omega}  c \, \d x
       - \int_{\Omega}  (\nabla \cdot \ubar) \,c\, \d x + \int_{\Omega}  \upri \cdot \nabla c \, \d x + \int_{\Gamma}  \gamma \, c \, \d x = 0 \,,
    \end{aligned}
\end{equation}
which suggests that for the quantity $c$ to be conserved and Eq.~(\ref{eq:weak-advec-decomp-play-2-w1}) to be identical to Eq.~(\ref{eq:weak-cons-w1}), the terms containing the velocity fields must vanish,
\begin{equation}\label{eq:mass-cons}
    \begin{aligned}
       \int_{\Omega}  \upri \cdot \nabla c \, \d x - \int_{\Omega}  (\nabla \cdot \ubar) \,c\, \d x = 0 \,.
    \end{aligned}
\end{equation}
More generally, the following must hold $\forall q \in H^1(\Omega)$,
\begin{equation}\label{eq:mass-cons-1}
    \begin{aligned}
       \int_{\Omega}  \upri \cdot \nabla q \, \d x = \int_{\Omega}  (\nabla \cdot \ubar) \,q\, \d x \,.
    \end{aligned}
\end{equation}
Eq.~(\ref{eq:mass-cons-1}) provides the key relation to ensure conservation of the developed scheme by allowing for the calculation of a field \(\upri\) for any given velocity field \(\ubar\), which may not necessarily be divergence-free. 

\noindent \textbf{Remark.}
\textit{
The condition Eq.~(\ref{eq:mass-cons-1}) means that $\b{u} = \ubar + \upri$ is weakly divergence-free,
\begin{equation}\label{eq:mass-cons-1a}
    \begin{aligned}
       \int_{\Omega} \nabla \cdot (\ubar + \upri) \, q \, \d x = 0 \,, \quad \forall q \in H^1(\Omega) \,.
    \end{aligned}
\end{equation}
To show that the two conditions~(\ref{eq:mass-cons-1}) and~(\ref{eq:mass-cons-1a}) are indeed equivalent, we write Eq.~(\ref{eq:mass-cons-1a}) as
\begin{equation}\label{eq:mass-cons-1b}
    \begin{aligned}
       \int_{\Omega} \nabla \cdot \ubar \, q \, \d x =  -\int_{\Omega} \nabla \cdot \upri \, q \, \d x \,, \quad \forall q \in H^1(\Omega) \,.
    \end{aligned}
\end{equation}
Integrating the right hand side by parts, and imposing $\upri = 0$ on $\Gamma$, results in Eq.~(\ref{eq:mass-cons-1}).
}
\qed

What remains is how to solve for $\upri$ given a $\ubar$. 
We define the fine scale velocity field $\upri$ to be the gradient of a scalar field $\varphi$, such that $\upri = \nabla \varphi$. 
Using this assumption in Eq.~(\ref{eq:mass-cons-1}) yields the weak form for the Poisson problem: find $\varphi \in \mathcal{Z} := \{ \, \phi \in H^1(\Omega): \int_{\Omega}\phi\, d\b{x} = 0 \, \}$ such that
\begin{equation}\label{eq:up-phi-laplace}
    \begin{aligned}
        \int_{\Omega}  \nabla q \cdot \nabla \varphi \, \d x = \int_{\Omega}   (\nabla \cdot \ubar) \,q\, \d x \,, \quad \forall q \in \mathcal{Z} \,.
    \end{aligned}
\end{equation}
The space $\mathcal{Z}$ for $\phi$ has been chosen so that the compatibility condition is satisfied for the given Poisson problem with the Neumann boundary condition $\grad \varphi \cdot \b{n} = 0 $ on $\Gamma$. 
Next, we outline the basics of the numerical implementation of the weak form using IGA.

\subsection{Numerical discretization and implementation}\label{sec:numerical}

For brevity, we focus on the implementation of the HW modified advective form~(\ref{eq:weak-advec-HWC}), as the advective form~(\ref{eq:weak-advec})  is a simpler version of the HW modified advective form, and the numerical implementation of the conservative form has already been given in~\cite{johnson_image-guided_2023}. 
For the discretization of the weak form~(\ref{eq:weak-advec-HWC}), we use a backward Euler scheme in time, and Galerkin method in space. 
We define $c^{\,n} \stackrel{\sim}{=} c(\b{x},t_n)$ where $t_n$ is a time instant $ t_0 = 0 < t_1 < t_2 < ... < t_k = T$ with uniform time interval $\Delta t$, such that the time-discretized weak form reads
\begin{equation}\label{eq:disc-time}
    \begin{aligned}
        \f{1}{\Delta t} \int_{\Omega} w \, (c^{n+1} - c^{\,n}) \, \d x &+ \int_{\Omega} D \, \nabla \, w \cdot \nabla c^{n+1} \, \d x + \int_{\Omega} w \, \ubar \cdot \nabla c^{n+1} \, \d x  + \int_{\Omega} w \, \upri \cdot \nabla c^{n+1} \, \d x \\
        &+ \int_{\Gamma} w \, \gamma \, c^{n+1} \, \d\Gamma    + \int_{\Omega} \sigma \, R(c^{n+1}) \, \ubar \cdot \nabla w \,\d x + \int_{\Omega} \sigma_p \, R_p(c^{n+1}) \, \upri \cdot \nabla w \,\d x = 0 \,.
    \end{aligned}
\end{equation}

\begin{figure}[h!]
    \centering
    \includegraphics[width=0.9\textwidth]{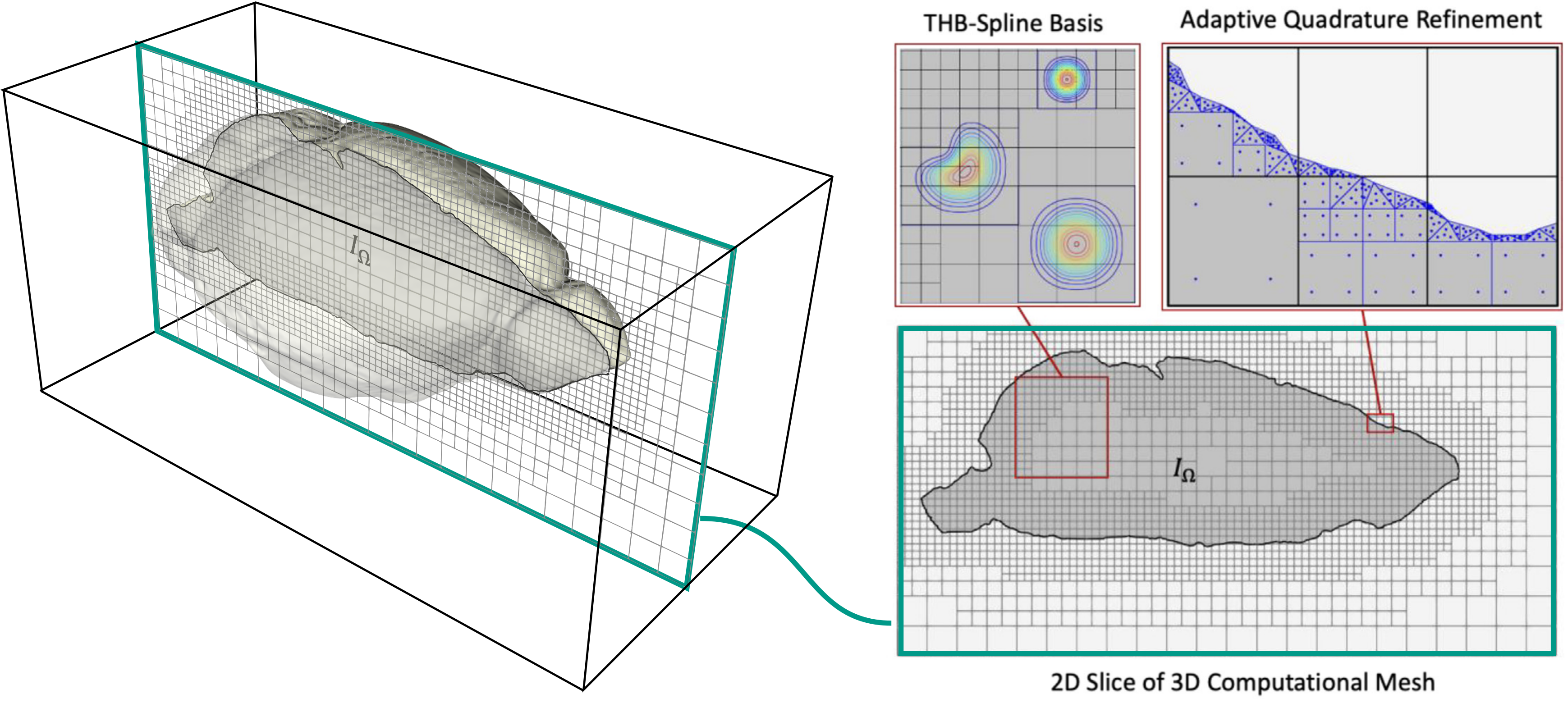}
    \caption{A schematic representation of a mouse brain geometry and meshing processes. A sample brain geometry is shown together with the immersed IGA mesh, truncated hierarchical B-spline (THB-spline) basis functions, and adaptive quadrature refinement~\cite{johnson_image-guided_2023}.}
    \label{fig:brain}
\end{figure}

For the discretization in space, we use an immersed finite element method~\cite{duster_finite_2008}.
Let $V$ denote the entire MR image volume. 
We create an extension of the problem from $\Omega$ to $V$, since $\Omega \subset V$, by introducing an indicator function $I_\Omega$ constructed directly from the brain volume segmentation, defined as
\begin{equation}
    I_\Omega = 
        \begin{cases}
        1, & \text{in } \Omega, \\
        0, & \text{in } V \setminus \Omega.
        \end{cases} \label{eq:indicator}
\end{equation}
The immersed formulation introduces discontinuities in the integrands, which are integrated using a quadrature rule with adaptive refinement on $\partial\Omega$, consisting of two steps: (i) elements intersecting $\partial\Omega$ are subdivided $n$ times using truncated hierarchical B-splines (THB splines)~\cite{giannelli_thb-splines_2016,buffa_complexity_2016}, and (ii) the resulting subelements are generated to accurately capture $\partial\Omega$, as illustrated in Fig.~\ref{fig:brain}.
This approach allows the discretization of $c$ to be defined on a mesh over $V$, avoiding the need to create boundary-conforming meshes in $\Omega$. 
This simplifies mesh generation and enables a smooth discretization over complex brain geometry.
Further details on the numerical implementation may be found in~\cite{johnson_image-guided_2023}.

The concentration $c$ is approximated in a finite-dimensional space $W^{\,h} \subset H^1(V)$ composed of quadratic THB-splines constructed on a hierarchically refined mesh on $V$. 
Basis functions with no support in $\Omega$ are removed from the discretization. 
Let $\{N_j(x)\}$ denote the resulting spline basis; then $c$ is written in terms of unknown coefficients $\{\alpha_j\}$ as
\begin{equation}
    c(\b{x},t_n) \approx c_h^n = \sum_j \alpha_j^n \, N_j(x)\,,\label{eq:ch-expansion}
\end{equation}
and the weighting function $w_h \approx w $ is taken to be $ N_i $.
Thus, the discretized weak form can be written as a system of equations 
\begin{equation}\label{eq:Wh-simp}
    \begin{aligned}
        & L_{ij} \, \alpha_j^{n+1}  = \Bigg[ \, M_{ij} + M^{\,s}_{ij} \, \Bigg]\, \alpha_j^{n}\, \quad \text{with} \quad L_{ij} := \Bigg[ \, M_{ij} + \Delta t \, \Big[ K_{ij} + \overline{A}_{ij} + A^{\prime}_{ij} + \, H_{ij} \, \Big] + S_{ij}  \, \Bigg] \,,
    \end{aligned}
\end{equation}
where the following arrays are defined as,
\begin{equation}\label{eq:Wh-ops}
    \begin{aligned}
        &M_{ij} := \, \intom N_i \, N_j \, \d x  \,,   \\
        &K_{ij}:= \intom D \, \nabla \, N_i \cdot \nabla N_j \, \d x \,  ,\\
        &\overline{A}_{ij} := \, \intom N_i \, (\ubar \cdot \nabla N_j) \, \d x \, ,\\
        &A^{\prime}_{ij}:= \intom N_i \, (\upri \cdot \nabla N_j) \, \d x \,, \\
        &H_{ij}:= \, \int_{\Gamma} \, \gamma \, N_i \, N_j \, \d\Gamma \,,\\
        &M^{\,s}_{ij} := \,\intom \, ( \sigma \ubar \cdot \nabla N_i) \,N_j \,\d x \,,   \\
        &S_{ij}:= M^{\,s}_{ij} + \Delta t \, \Bigg[ \, \intom  (\sigma \,\ubar \cdot \nabla N_i)\,\Big[\, \ubar \cdot \nabla N_j - \nabla \cdot (D \nabla N_j)\, \Big] \,\d x + \intom  \, (\sigma_p \, \upri \cdot \nabla N_i) \, (\upri \cdot \nabla N_j)\,\d x \, \Bigg] \,.
    \end{aligned}
\end{equation}
Assuming we know $\ubar$, we can solve for $\varphi$ in Eq.~(\ref{eq:up-phi-laplace})  to find the velocity field component $\upri$.
We discretize $ q^{\,h} \approx q $ by $N_i$ and $ \varphi^{\,h} \approx \varphi $ by $ \sum_j \varphi_j^{\,h} \, N_j $, so that $\nabla \varphi^{\,h} = \sum_j \varphi^{\,h}_j \, \nabla N_j$, to obtain the system
\begin{equation}\label{eq:up-phi-laplace-sys}
    \begin{aligned}
        \int_{\Omega}  \nabla N_i \cdot \nabla N_j \, \d x \, \varphi^{\,h}_j = \int_{\Omega}   (\nabla \cdot \ubar) \,N_i\, \d x \,.
    \end{aligned}
\end{equation}
Eq.~(\ref{eq:up-phi-laplace-sys}) can also be written as the linear system
\begin{equation}\label{eq:up-phi-laplace-sys-1}
    \begin{aligned}
        \kappa_{ij} \, \varphi^{\,h}_j = f_i\,,
    \end{aligned}
\end{equation}
where
\begin{equation}\label{eq:up-phi-laplace-sys-defs}
    \begin{aligned}
        \kappa_{ij} &:= \int_{\Omega}  \nabla N_i \cdot \nabla N_j \, \d x\,, \\ 
        f_i &:= \int_{\Omega}   (\nabla \cdot \ubar) \,N_i\, \d x \,.
    \end{aligned}
\end{equation}
Thus, given a field $\ubar$, the corresponding field  $\upri = \nabla \varphi^{\,h} = \sum_j \varphi^{\,h}_j \, \nabla N_j$ can be found by solving the linear system in Eq.~(\ref{eq:up-phi-laplace-sys-1}) for $\varphi^{\,h}$.
The velocity field $\ubar$ would come from one of a variety of sources such as the direct output of experimental data, the output of numerical simulations with the Navier-Stokes equations (in which case $\ubar$ may not be pointwise solenoidal) or 
by solving the inverse problem for a set of experimental data. 
Section~\ref{sec:inverse} deals with this question and proposes an inverse modeling framework to solve for $\ubar$, along with the other unknown parameters in the model.

\section{Reconstruction of a divergence-free velocity field from image data}\label{sec:inverse}

This section formulates the inverse problem of reconstructing a divergence-free velocity field from image data.
Section~\ref{sec:exp} details the experimental image data utilized in this work and the various constraints inherent to these datasets.
Section~\ref{sec:invform} presents the mathematical framework for solving for the velocity field, along with the other unknown transport fields (diffusion and clearance), within the previously established finite element environment. Finally,
Section~\ref{sec:numimp} describes the numerical implementation of the inverse problem.

\subsection{Experimental image data and inverse model definition}\label{sec:exp}

\begin{figure}[h!]
    \centering
    \includegraphics[width=0.9\textwidth]{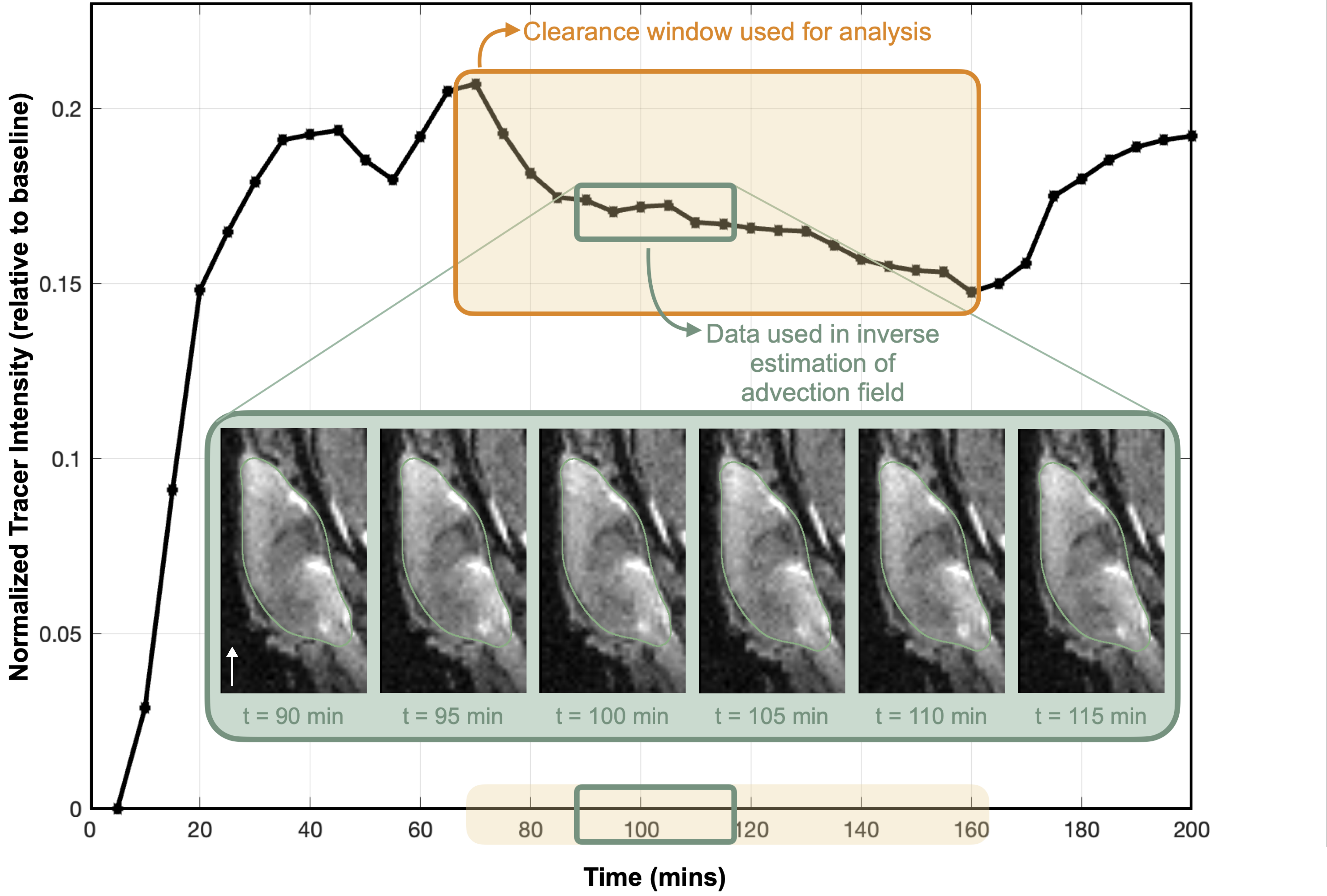}
    \caption{Normalized total tracer intensity vs. time at each imaging instance of the experiment. The orange box highlights data relevant to the glymphatic clearance window. The green box highlights the calibration dataset used in the inverse approach.}
    \label{fig:experiment}
\end{figure}

In our previous work~\cite{johnson_image-guided_2023}, the diffusivity field was derived from the Apparent Diffusion Coefficient (ADC) map obtained via Diffusion-Weighted MRI (DW-MRI) of the brain. 
Furthermore, the clearance parameter was set to  zero on the cranial boundary and non-zero elsewhere.
 Due to the lack of CE-MRI data, the advective field in~\cite{johnson_image-guided_2023} was also derived from available DW-MRI data.
In this study, we infer these unknown fields, in particular the advective field, using CE-MRI data obtained from our own tracer experiment. 
In the experiment, a soluble tracer is injected into the cerebrospinal fluid (CSF) pathways of the mouse brain through the cisterna magna. 
The tracer appears at an intensity higher than that of the surrounding tissue in MR images, allowing its distribution to be tracked over time as it travels through the CSF and is cleared from the brain, acting as a surrogate for metabolic waste products. 
The subject is imaged before tracer injection and then at five-minute intervals over several hours post-injection.
This produces a pre--contrast baseline image and a sequence of post-contrast images that captures the dynamic transport and clearance of the tracer through the glymphatic system.
Fig.~\ref{fig:experiment} plots the normalized total tracer intensity at each imaging instance throughout the experiment.
The tracer is injected at time $t=5$ mins, producing an initial spike in total signal as the tracer enters and rapidly fills fluid-filled areas of the brain.
Representing the data in this form makes it possible to identify the time window relevant to glymphatic clearance sufficiently removed from the perturbations to the system caused by the injection.
Glymphatic clearance occurs in the highlighted orange window, where a clear decline in signal intensity is observed.
A subset of the data points in this region, indicated by the green box, is initially selected for calibration of the unknown parameters, while the remaining points in the orange window are used for validation.
A formal sensitivity analysis regarding the optimal subset and number of calibration points is beyond the scope of the present study and remains a subject for future investigation. 
In this work, a five-step calibration window was selected to ensure that the model  (i) captures a sufficient range of clearance without having to deal with overfitting, and (ii) does not rely on too few data points that it only captures transient or noise-induced behavior.

\subsubsection{A finite difference approach}
\begin{figure}[h!]
    \centering
    \includegraphics[width=0.99\textwidth]{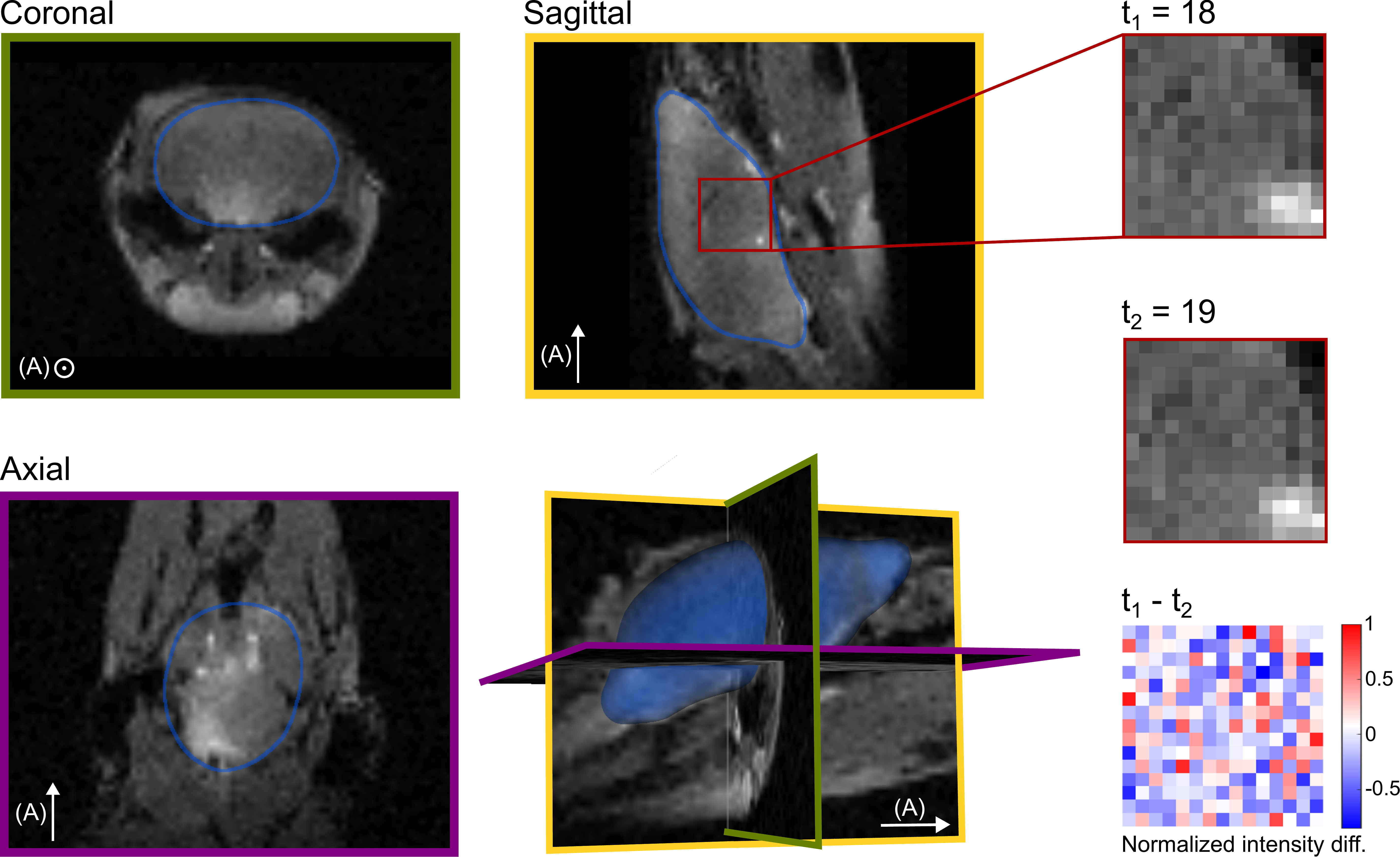}
    \caption{(Left panel) CE-MRI slices (axial, sagittal, coronal) with the domain of interest shown in blue. 
    (Right panel) The zoomed regions at two consecutive time points illustrate the very small intensity changes compared to the underlying noise and coarse resolution. 
    The normalized $t_1 - t_2$ map highlights the noise in the data, showing the limitations of using these image data directly for velocity estimation.}
    \label{fig:MRI}
\end{figure}

A common approach for solving the governing PDE in Eq.~(\ref{eq:strong}) is to discretize it directly using a finite difference scheme. 
After discretization, the resulting operators can be inverted to obtain a velocity field from the image data.
The derivatives in Eq.~(\ref{eq:strong}) are discretized using central differences in time and space, where the step size for time is the imaging time interval, $\Delta t = 5 $ minutes, and the step size in space is the voxel size of the images, $\Delta x, \Delta y, \Delta z = 0.3$ mm.
Constructing the finite difference scheme yields a set of equations for each voxel that can be solved for each of the velocity field components.
Although straightforward, this approach poses significant challenges when used with noisy, real-world imaging data. 
Fig.~\ref{fig:MRI} illustrates the CE-MRI data from a single time point (in coronal, sagittal, and axial directions), where the blue region indicates the simulation domain $\Omega$ of the brain.
In this setting, the velocity is approximated to be proportional to the temporal gradient of the tracer concentration, calculated as the difference in concentration over a discrete timestep. 
However, as shown in the right inset of Fig.~\ref{fig:MRI}, the inherent noise and discretization errors in the imaging data render this voxel-wise differencing procedure unreliable.
The resulting checkerboard plot reveals a non-physical, high-frequency noise distribution in the velocity surrogate, characterized by a lack of spatial coherence at the voxel level.

\begin{figure}[h!]
    \centering
    \includegraphics[width=0.99\textwidth]{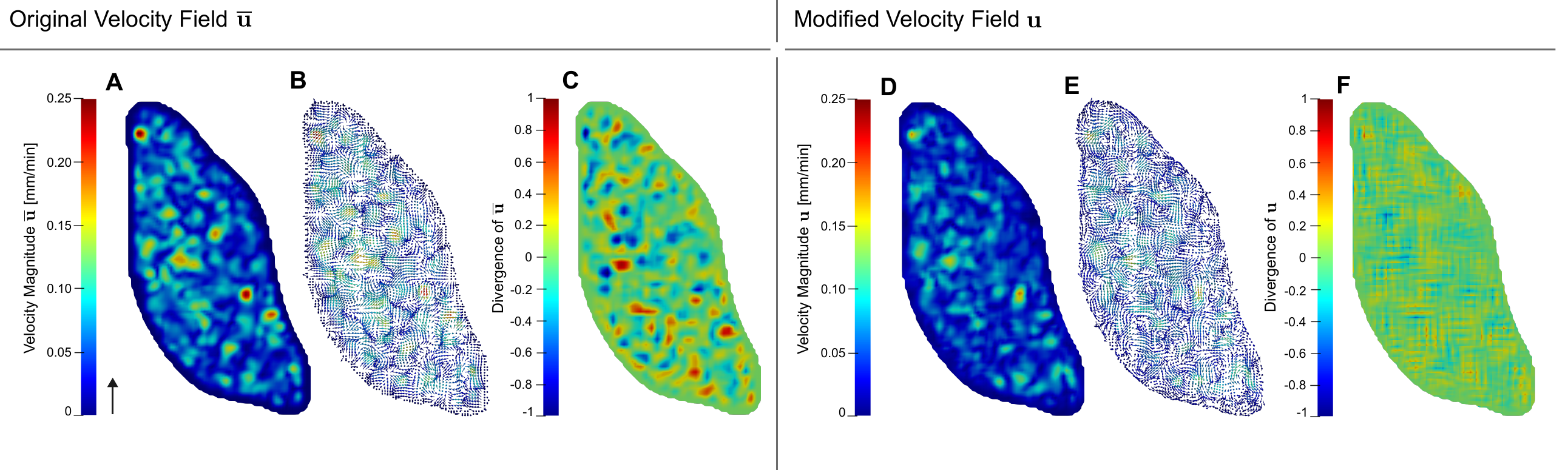}
    \caption{The velocity field obtained through finite differencing is shown here projected onto the brain geometry and the corresponding B-spline mesh on a sagittal slice (black arrow indicates the anterior direction). The left half (A–C) shows the original field \(\ubar\): the velocity magnitude in (A) contains scattered peaks likely originating from MRI noise and resolution limits, and the corresponding vector field in (B) is clearly not divergence-free, with locally irregular directions, as further illustrated by its divergence map in (C), which portrays non-physical sources and sinks throughout the domain. The right half (D–F) shows the modified velocity \(\b{u}=\ubar+\upri\): (D) the magnitude, (E) the vector field, and (F) the divergence map.}
    \label{fig:ubar}
\end{figure}

The HW modified approach introduced in Section~\ref{sec:gen} allows a velocity field obtained through finite differencing, despite the shortcomings described above, to be used in forward simulations of glymphatic transport within a concentration-conserving advection–diffusion scheme. 
The raw velocity field extracted from the image data using finite differences, $\ubar$, and its modification under the current approach, $\ubar + \upri$, are presented in Fig.~\ref{fig:ubar}, in insets (A-C) and (D-F), respectively.
In constructing the system of equations to solve for the advection unknowns using finite differencing, and in the following forward simulations whose results are given in Section~\ref{sec:resultsFD}, the diffusivity and clearance parameters are consistent with those in~\cite{johnson_image-guided_2023}, with one key modification: the diffusivity map derived from the ADC values has been scaled by a factor of $0.3$. 
This scaling reflects a Stokes-Einstein correction accounting for the molecular weight and size differences between water molecules, whose diffusivity is reported in the ADC of DW-MRI, and the tracer molecules used in the experiments. 
The velocity field obtained from finite differencing exhibits randomly distributed points of high velocity magnitude, as seen in Fig.~\ref{fig:ubar}A. 
Furthermore, the vector field in Fig.~\ref{fig:ubar}B and the divergence of said vector field in Fig.~\ref{fig:ubar}C, reveal that the field generated is not divergence free. 
Contrary to the incompressibility expected of the flow, the generated velocity field contains multiple spots of significant positive and negative divergence, which when used in the conservative form of the advection-diffusion equation, results in spurious, non-physical sinks and sources across the domain.
Fig.~\ref{fig:ubar}~(D-F) shows the HW modified velocity field $\b{u} = \ubar + \upri$ where the velocity field $\ubar$ from inset A has been used in~(\ref{eq:up-phi-laplace-sys}) to obtain $\upri$. 
Fig.~\ref{fig:ubar}D reveals a much smoother field compared to Fig.~\ref{fig:ubar}A due to the corrective approach. 
The vector map in Fig.~\ref{fig:ubar}E and divergence map in Fig.~\ref{fig:ubar}F further confirm that the advection field no longer contains the large sinks and sources introduced by the non-zero divergence of the original velocity field, and that $\b{u} = \ubar + \upri$ is weakly divergence free.

These findings from Fig.~\ref{fig:ubar} makes it critical to find a novel strategy to infer $\ubar$ without succumbing to the shortcomings caused by image noise and discretization. 
This motivates embedding the HW modified approach directly within the inverse problem, where it can replace the finite-difference-type velocity estimation  of $\ubar$ from image data. 
Section~\ref{sec:invform} formulates this inverse problem of recovering the unknown model parameters from the available data on a finite element framework using the HW modified approach.

\subsection{Formulation of the inverse problem}\label{sec:invform}
The weak-form finite element formulation established for the forward problem provides a natural basis for the full inverse problem. This variational framework enables the simultaneous recovery of the velocity field $\ubar$, the diffusivity field $D$ and the boundary clearance parameter $\gamma$ directly from the available spatiotemporal data.
Thus, the diffusive and advective fields, $D$ and $\ubar$, along with the boundary clearance distribution $\gamma$, become the unknowns in the inverse problem.
We can discretize the three unknown fields the same way we discretized concentration $c$ in Eq.~(\ref{eq:ch-expansion}),
\begin{equation}\label{eq:disc-inv-params}
    \begin{aligned}
       &D &&\approx D^{\,h} = \sum_{j=1}^{n_{cp}} D_j^{\,h} \, N_j \,, \\
       & \ubar &&\approx \overline{u}^h_k = \sum_{j=1}^{n_{cp}} \overline{u}_{jk}^h \, N_j\, \quad \text{where} \quad \overline{u}_k = \ubar \cdot \mathbf{e}_k,\\
       &\gamma &&\approx \gamma^{\,h} = \sum_{j=1}^{n_{cp}} \gamma^{\,h}_j \, N_j \,,
    \end{aligned}
\end{equation}
where $\mathbf{e}_k$ are the Cartesian unit vectors in $\mathbb{R}^3$.
The basis functions $N_j$ are the same basis functions used for the spatial discretization of $c$, and $n_{cp}$ is the number of control points in the discretization.
For convenience, we introduce a vector-valued basis $\b{N}^*$
constructed from the scalar basis functions,
\begin{equation}
\b{N}^*_{jk}
:=
N_j \,\mathbf{e}_k,
\qquad
j=1,\dots,n_{cp},\;\; k=1,2,3.
\end{equation}
This allows us to write the velocity components in a flattened fashion such that 
\begin{equation}
\bar{\mathbf{u}}_{\mathrm{vec}}^h
=
\begin{bmatrix}
\bar{u}_{11}^h &
\cdots &
\bar{u}_{n_{cp}1}^h &
\bar{u}_{12}^h &
\cdots &
\bar{u}_{n_{cp}2}^h &
\bar{u}_{13}^h &
\cdots &
\bar{u}_{n_{cp}3}^h \,
\end{bmatrix}^T
\in \mathbb{R}^{3n_{cp}} \,, 
\end{equation}
so that the velocity field can be written compactly as
\begin{equation}
\ubar \approx \overline{u}_k^h
=
\sum_{m=1}^{3n_{cp}}
\{ \bar{u}_{\mathrm{vec}}^h \}_m \,N_{mk}^*.
\end{equation}
We can then define a vector of unknowns $\b{\Theta}$ to collect all the unknown coefficients,
\begin{equation}\label{eq:theta}
    \begin{aligned}
\b{\Theta}
:=
\begin{bmatrix}
\mathbf{D}^h\\
\bar{\mathbf{u}}_{\mathrm{vec}}^h\\
\boldsymbol{\gamma}^h
\end{bmatrix}
\qquad
\text{where}
\qquad
\begin{matrix}
\mathbf{D}^h:=\{ D_1^h,\ldots,D_{n_{cp}}^h \}^t,\; \\
\bar{\mathbf{u}}_{\mathrm{vec}}^h:=\{ u_{11}^h,\ldots,u_{n_{cp}3}^h\}^t,\;\\
\boldsymbol{\gamma}^h:=\{ \gamma_1^h,\ldots,\gamma_{n_{cp}}^h\}^t.
\end{matrix}
    \end{aligned}
\end{equation}
Let $c_{\b{\Theta}}^{\,n} := c \, (\,\b{x},t^{\,n},\b{\Theta}\,)$ denote the concentration of tracer at point $\b{x} \in \Omega$, time $t^{\,n}$, as a function of the parameter set $\b{\Theta}$ and satisfying Eq.~(\ref{eq:weak-advec-HWC}). 
Furthermore, let $\hat{c}^{\,n}$ denote the CE-MRI data at time $t^{\,n}$. 
The minimization problem is then finding the optimal set of parameters ${\b{\Theta}}^{\,*}$ such that the total difference between the computed and experimental concentration distributions at timestep $t^k$ where $t \in \{ \,t^1,...,t^n,...,t^k\,\}$ is minimized. 
In other words, we want to find $\b{\Theta}^{\,*}$ such that 
\begin{equation}\label{eq:min-J}
    \begin{aligned}
        \b{\Theta}^{\,*} = \text{argmin} \, J ^{k}( \b{\Theta}) \quad \text{where} \quad J^{k}( \b{\Theta}) := \f{1}{2} \, \int_{\Omega} \, \Big[\, c^{\,k}_{\b{\Theta}} - \hat{c}^{\,k} \,\Big]^2 \, \d x \,.
    \end{aligned}
\end{equation}
We solve Eq.~(\ref{eq:min-J}) by linearizing $c_{\b{\Theta}}^{\,k}$ about some point $\b{\Theta}^{\,\ell}$ where $\ell$ is the iteration number,
\begin{equation}\label{eq:lin-c}
    \begin{aligned}
        c^{\,k}_{\b{\Theta}} \approx c^{\,k}_{\b{\Theta}^{\,\ell}} + \f{\p c^{\,k}_{\b{\Theta}}}{\p \, \b{\Theta}} \Bigg|_{\b{\Theta}=\b{\b{\Theta}^{\,\ell}}} \cdot \Delta \b{\Theta} \quad \text{where} \quad \b{\Theta}^{\ell+1} = \b{\Theta}^{\ell} + \Delta \b{\Theta}\,,
    \end{aligned}   
\end{equation}
so that the minimization problem becomes finding $\b{\Theta}^{\,*}$ such that
\begin{equation}\label{eq:min-J-c}
    \begin{aligned}
        \b{\Theta}^{\,*} = \text{argmin} \, J^k ( \b{\Theta}) \quad \text{where} \quad J^{k}( \b{\Theta}) = \f{1}{2} \, \int_{\Omega} \, \Bigg[\, c^{\,k}_{\b{\Theta}^{\,\ell}} + \f{\p c^{\,k}_{\b{\Theta}}}{\p \, \b{\Theta}} \Bigg|_{\b{\Theta}=\b{\b{\Theta}^{\,\ell}}} \cdot \Delta \b{\Theta} \, - \, \hat{c}^{\,k} \,\Bigg]^{\,2} \, \d x\,.
    \end{aligned}
\end{equation}
The functional $J^{k}( \b{\Theta})$ as given in Eq.~(\ref{eq:min-J-c}) attains its minimum when
\begin{equation}\label{eq:min-J-c-zero}
    \begin{aligned}
        \f{\p J^{k} ( \b{\Theta})}{\p \Delta \b{\Theta}} = 0 \quad \Rightarrow \quad 
        \int_{\Omega} \, \Bigg[\, c^{\,k}_{\b{\Theta}^{\,\ell}} + \f{\p c^{\,k}_{\b{\Theta}}}{{\p \, \b{\Theta}}} \Bigg|_{\b{\Theta}=\b{\b{\Theta}^{\,\ell}}} \cdot \Delta \b{\Theta} - \hat{c}^{\,k} \,\Bigg] \cdot \, \f{\p c^{\,k}_{\b{\Theta}}}{{\p \, \b{\Theta}}}\Bigg|_{\b{\Theta} = \b{\b{\Theta}^{\,\ell}}} = 0 \,,
    \end{aligned}
\end{equation}
which when rearranged, yields the system
\begin{equation}\label{eq:min-J-c-zero-sys}
    \begin{aligned}
        \int_{\Omega} \, \Bigg[\, \f{\p c^{\,k}_{\b{\Theta}}}{{\p \, \b{\Theta}}} \, \dyad \f{\p c^{\,k}_{\b{\Theta}}}{\p \, \b{\Theta}} \, \,\Bigg] \Bigg|_{\b{\Theta} = \b{\Theta}^{\,\ell}} \, \d x \cdot \Delta \b{\Theta}  = \int_{\Omega} \, \Bigg[\, \hat{c}^{\,t} - c^{\,k}_{\b{\Theta}^{\,\ell}} \,\Bigg] \cdot \, \f{\p c^{\,k}_{\b{\Theta}}}{{\p \, \b{\Theta}}}\Bigg|_{\b{\Theta} = \b{\Theta}^{\,\ell}}\,\d x\,,
    \end{aligned}
\end{equation}
where $\Delta \b{\Theta}$ is the unknown and needs to be solved for, and the remaining quantities can be derived from the forward problem being solved until timestep $k$ and by differentiating the forward problem with respect to the input parameters $\b{\Theta}$.
The change of the concentration field at the final timestep $k$ with respect to a change in the parameters $\b{\Theta}$ is
\begin{equation}\label{eq:dcdT}
    \begin{aligned}
        \f{\p \, c^{\,k}_{\b{\Theta}}}{{\p \, \b{\Theta}}} \approx \f{\p \, (c^{\,k}_{\b{\Theta}})^{\,h}}{{\p \, \b{\Theta}}} = \sum_{j}\f{\p \, \alpha_j^{\,k}}{\p \, \b{\Theta} } \, N_j  \,.
    \end{aligned}
\end{equation}
To obtain an expression for Eq.~(\ref{eq:dcdT}), we differentiate the discretized system of the forward problem~(\ref{eq:Wh-simp}) with respect to $\b{\Theta}$,
\begin{equation}\label{eq:Wh-simp-diff}
    \begin{aligned}
       & L_{ij} \, \f{\p \, \alpha_j^{\,n+1}}{\p \, \b{\Theta}}  = \Big[ \,M_{ij} + M^{\,s}_{ij} \, \Big] \, \f{\p \, \alpha_j^{\,n}}{\p \, \b{\Theta}} \, - \, \f{\p L_{ij}}{\p \, \b{\Theta}} \, \alpha^{\, n+1}_j \,,
    \end{aligned}
\end{equation}
where $L_{ij}$, $M_{ij}$, and $M^{\,s}_{ij}$ have been defined in Eq.~(\ref{eq:Wh-ops}) and are independent of the solution $\alpha_j$. 
For simplicity, we have assumed that the stabilization term $S_{ij}$ is explicit on iterations of the inverse problem, that is $S_{ij} := S_{ij}(\b{\Theta}^{\ell-1})$. 
Defining the stabilization terms to be functions of the set of parameters from the previous iteration in this way allows the derivatives of the stabilization terms with respect to the parameters $\b{\Theta}$ to disappear, with $\f{\p S_{ij}(\b{\Theta}^{\ell-1})}{\p \b{\Theta}} = 0$.
Note that $\b{L}$ and $\b{M}$ are of size $n_{cp} \times n_{cp}$, and $\f{\p \,\alpha_j^{\,n+1}}{\p \, \b{\Theta}}$ is of size $n_{cp} \times n_{\Theta}$, where $n_{\Theta}$ is the number of elements in $\b{\Theta}$.
The detailed expression for $\f{\p L_{ij}}{\p \, \b{\Theta}} \, \alpha^{\, n+1}_j$ obtained by differentiating the discretized weak form of the forward problem can be found in~\ref{sec:app}.
Thus, Eq.~(\ref{eq:Wh-simp-diff}) presents a linear system to be solved for $\f{\p \, \alpha_j^{n+1}}{\p \, \b{\Theta}}$ at each timestep of the forward problem.

\subsection{Numerical implementation}\label{sec:numimp}

\begin{figure}[h!]
    \centering
    \includegraphics[width=0.94\textwidth]{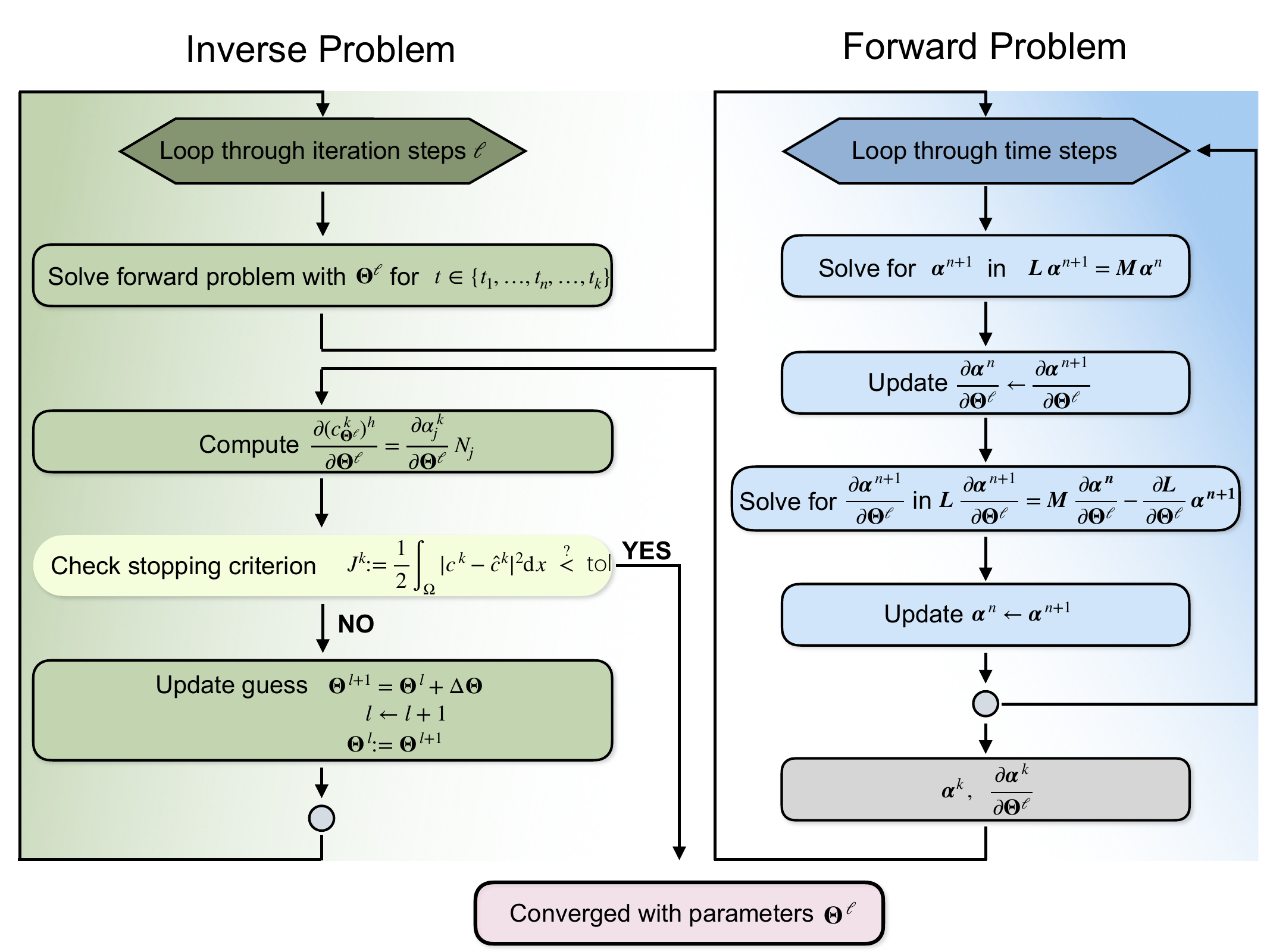}
    \caption{Flowchart for the inverse problem of solving for the unknown fields in the glymphatic transport problem.}
    \label{fig:flow}
\end{figure}

Fig.~\ref{fig:flow} provides a flowchart that outlines the solution algorithm for the inverse problem.
For each iteration $\ell$ over the parameters $\b{\Theta}$, we solve the forward system with inputs $\b{\Theta}$ for $t \in \{t_1, \dots, t_n, \dots, t_k\}$, while solving the system in Eq.~(\ref{eq:Wh-simp-diff}) at each timestep. 
Solving for $\f{\p \alpha_j^{\,k}}{\p \Theta_m}$ at the final timestep $k$ yields $\f{\p \, (c^{\,k}_{\b{\Theta}})^{\,h}}{{\p \, \b{\Theta}}}$ via Eq.~(\ref{eq:dcdT}). 
We can then solve for $\Delta \b{\Theta}$ in Eq.~(\ref{eq:min-J-c-zero-sys}), which is the update for the parameters for the next step of iteration.
Note that evaluating the residual $J^k$ requires projecting the image data onto the same mesh used for the simulation. 
This is an additional benefit of the approach adopted here, since projecting the image data onto the B-spline basis provides a natural smoothing effect and substantially reduces the noise in the image data, while simultaneously ensuring the smoothness of the parameter fields. 

The inverse problem is solved using a Newton-Raphson scheme implemented in Python using the Nutils library~\cite{nutils9}. 
We employ quadratic THB-splines, which provide implicit regularization of the image data due to their $C^1$-continuity, filtering noise. 
Simulations are performed with a fixed time step of $\Delta t = 5$ minutes.
Time accuracy was checked by systematically refining this time step size. 
Results presented herein represent converged results with respect to time step size. 
Given that the expected residual value is dependent on the inherent noise of the experimental CE-MRI data, convergence of the inverse algorithm is not defined by a fixed tolerance; instead, the solver terminates when the objective function reaches a stable plateau, indicating that a local numerical minimum has been attained.
The inverse problem is solved on a mesh of initial size [1.0, 1.0, 1.0] mm, which goes through one level of mesh refinement, and one level of quadrature refinement at the boundaries. 
The resultant fields of the inverse problem are then used in the forward problem which is solved on a finer mesh of [0.5, 0.5, 0.5] mm, with one level of uniform mesh refinement, and two levels of quadrature refinement at the boundaries.
The inverse problem is intentionally solved on a  coarser mesh to manage the size of the linear systems solved at each Newton iteration.
The dimensions of the linear systems in the Newton iterations scale rapidly with mesh refinement, significantly increasing the computational cost per iteration, because the unknowns are the B-spline coefficients of the reconstructed fields. 
Using a coarser mesh allows for a faster convergence to the numerical minimum while still capturing the essential features of the velocity field.
Even with the coarser mesh, the $n_{cp} \approx 5000 $, which makes the number of unknown parameters inferred $5 \times n_{cp} \approx 25000$.

\section{Numerical results and discussion}\label{sec:resultsgen}

This section presents numerical results demonstrating the performance of the proposed framework in reconstructing glymphatic velocity fields. 
In Section~\ref{sec:resultsFD}, we demonstrate the utility of the HW-modified approach by performing forward transport simulations using velocity fields derived directly from experimental data using finite differences.
Subsequently, Section~\ref{sec:invres} presents the ability of the Newton-based inverse algorithm to recover physically consistent velocity, diffusivity, and clearance parameters from spatiotemporal MRI data. 

\subsection{Forward simulations using finite-difference-derived velocity fields}\label{sec:resultsFD}

This section presents the results of forward simulations of glymphatic transport with the three different methods outlined in Sec.~\ref{sec:gen}, using the flawed finite-difference velocity fields illustrated in Fig.~\ref{fig:ubar}.
The clearance conditions follow those in~\cite{johnson_image-guided_2023}, while the diffusivity is derived from ADC values scaled by a factor of 0.3 via a Stokes-Einstein relation to account for tracer-specific molecular properties.
The initial condition for the forward simulation is the tracer intensity distribution at time $t=70$ minutes of the experiment, projected onto the same mesh.
The forward problem is run for 120 minutes, with 5 minute time intervals. 

\begin{figure}[h!]
    \centering
    \includegraphics[width=0.95\textwidth]{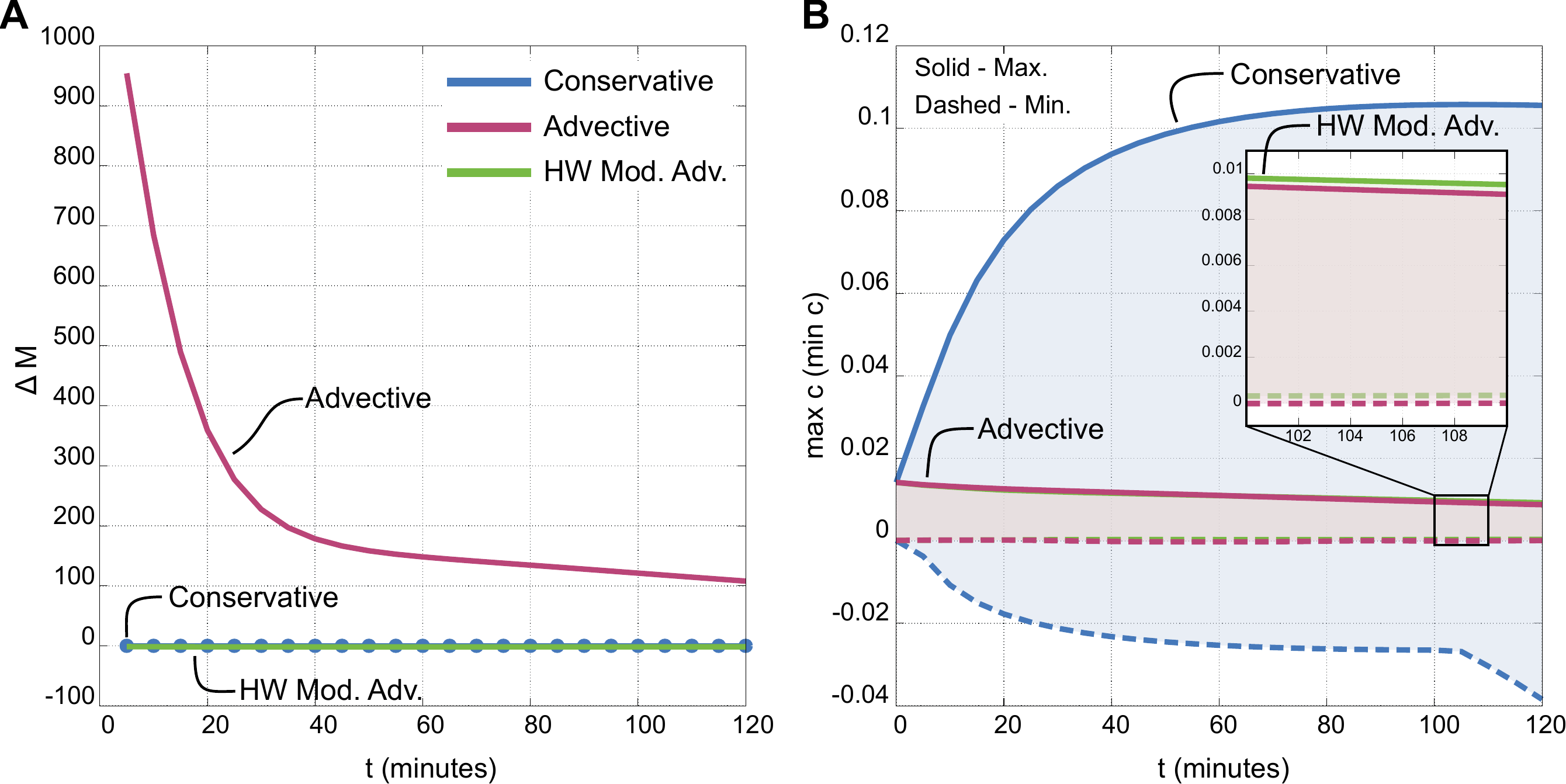}
    \caption{Calculated values for conservation using Eq.~(\ref{eq:deltaM}) vs. time (inset A) and maximum and minimum concentration over the whole domain vs. time (inset B). 
    Inset A shows that the advective form does not conserve $c$, while the conservative and modified advective forms do. 
    Inset B shows that the conservative form causes positive and negative instabilities in the concentration field.}
    \label{fig:graph}
\end{figure}

Fig.~\ref{fig:graph}A plots a conservation parameter $\Delta M$ over simulation time for the conservative, advective, and HW modified advective methods.
We define the parameter $\Delta M$ for the conservation of the quantity $c$ to be
\begin{equation}\label{eq:deltaM}
    \begin{aligned}
     \Delta M(t_{n}) := \lb \int_{\Omega} c_h^{\,n-1}(\b{x})\, \d x - \int_{\Omega} c_h^{\,n}(\b{x})\, \d x \rb
        - \Delta t \int_{\Gamma} \gamma\, c_h^{\,n}(\b{x}) \, dS \,.
\end{aligned}
\end{equation} 
This parameter~(\ref{eq:deltaM}) is essentially the time and space discretized version of the fundamental condition for conservation given in Eq.~(\ref{eq:weak-cons-w1}).
Note that the conservation parameter $\Delta M$ should be zero at all time points.
Fig.~\ref{fig:graph}A shows that $\Delta M \neq 0$ for the advective form, proving that it does not conserve $c$, whereas the conservative and HW modified advective forms do.
Fig.~\ref{fig:graph}B displays the maximum (solid) and minimum (dashed) concentration values over time for the three methods.
Since there is no source term, and we start with an initial load of tracer, the maximum concentration is expected to decrease monotonically, until the point where all of it is cleared from the domain at a sufficiently large time. 
Similarly, since concentration is a strictly positive quantity, the minimum concentration over the domain over all time steps is expected to be greater than or equal to zero. 
The advective and HW modified advective forms display these physically correct behaviors. 
For the conservative form which uses the unmodified velocity field $\ubar$, this plot quantitatively confirms that positive and negative divergences occur, revealing its unstable nature as simulation time increases.
Running the forward simulations for a longer time, until $t=5760$ minutes, the maximum concentration for the advective and HW modified advective forms reaches zero, as expected.
However, the conservative form yields unstable results, with the final concentration values over the domain at time $t=5760$ minutes ranging between $-1.27\times10^{42}$ and $2.67\times10^{39}$, clearly indicative of blow-up, as suggested by earlier timesteps in Fig.~\ref{fig:graph}B.

\begin{figure}[h!]
    \centering
    \includegraphics[width=0.9\textwidth]{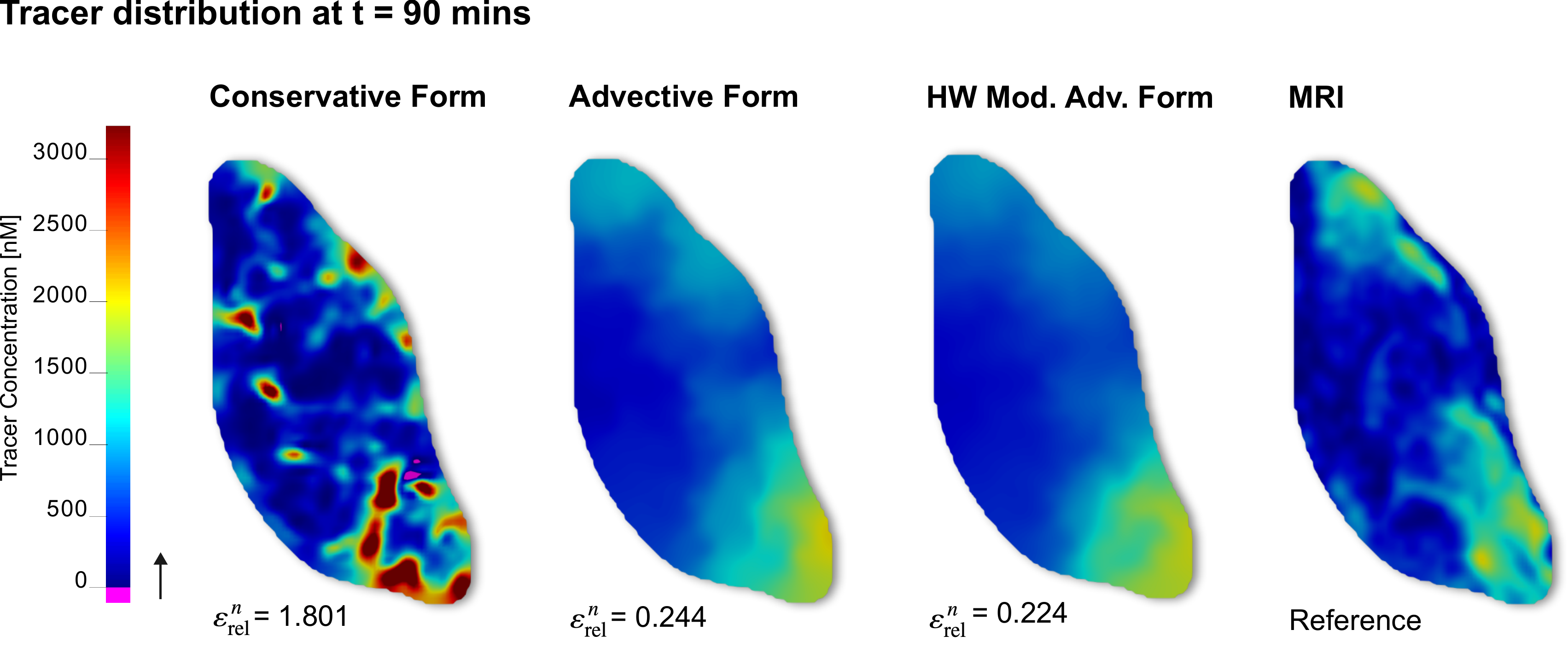}
    \caption{Simulation results for the three methods at time $t=90$ minutes, alongside CE-MRI data projected onto the computational mesh at the same time point. Arrow indicates anterior direction. Quantities at the bottom show relative error between simulated tracer concentration and image tracer concentration at $t=90$ minutes, calculated by~(\ref{eq:J-rel}).}
    \label{fig:compresults}
\end{figure}

The agreement of the results of the three methods using the finite difference derived velocity field with experimental data is shown in Fig.~\ref{fig:compresults}.
The simulation results for the three methods is displayed at time $t=90$ minutes, together with the segmented CE-MRI data projected onto the computational mesh at the same time. 
The relative error between the predicted concentration distribution $c^{\,n}$ at timestep $n$, and expected concentration distribution (CE-MRI) $\hat{c}^{\,n}$ is shown below each figure, where the relative error is defined as 
\begin{equation}\label{eq:J-rel}
    \begin{aligned}
        \varepsilon_{\text{rel}}^{\,n} = \, \f{\int_{\Omega} \, \Big[\, c^{\,n}_{h} - \hat{c}^{\,n} \,\Big]^2 \, \d x}{\int_{\Omega} \, \Big[\hat{c}^{\,n} \,\Big]^2 \, \d x} \,.
    \end{aligned}
\end{equation}
The error is highest for the conservative form, which again exhibits some negative areas of concentration and pronounced areas of accumulation (dark red areas). 
This is expected because even though the method conserves the quantity $c$, the divergence of the uncorrected velocity field $\ubar$ creates sinks and sources in the domain. 
The HW modified advective form provides the closest agreement with the experimental data among the three methods. 
This finding is significant since it shows that the HW modified advective form works well even when a velocity field obtained through finite-differences is taken as the experimentally obtained velocity field $\ubar$.

In summary, the conservative form results in a scheme that conserves $c$, but is unstable in the long run due to its spurious sinks and sources in the domain, whereas the advective form provides long-term stability, but lacks conservation. 
We have demonstrated that the HW modified advective form ensures conservation, and  yields the most accurate and physically consistent results for using image-derived velocity fields in solving the advection diffusion equation.
The three approaches are summarized in Table~\ref{tab:comp}. 

\renewcommand{\arraystretch}{1.0}
\begin{table}[h!]
    \centering
    \small
    \begin{tabular}{
        >{\raggedright\arraybackslash}p{4cm}
        >{\raggedright\arraybackslash}p{4cm}
        >{\centering\arraybackslash}p{3cm}
        >{\centering\arraybackslash}p{3cm}
    }
    \toprule
    \textbf{Method} &
    \textbf{Advective Term} &
    \textbf{Conservation} &
    \textbf{Long-Term Stability} \\
    \midrule
    \textbf{Conservative} &
    $\displaystyle \int_{\Omega} w\, \nabla \!\cdot (c \b{u})\, \d x$ &
    \cmark & \xmark \\
    
    \addlinespace[6pt]
    \textbf{Advective} &
    $\displaystyle \int_{\Omega} w\, \b{u}\!\cdot\!\nabla c\, \d x$ &
    \xmark & \cmark \\
    
    \addlinespace[6pt]
    \textbf{HW Mod. Advective} &
    $\displaystyle 
    \int_{\Omega} w\, \bar{\b{u}}\!\cdot\!\nabla c\, \d x
    + \int_{\Omega} w\, \b{u}'\!\cdot\!\nabla c\, \d x$ &
    \cmark & \cmark \\
    \bottomrule
    \end{tabular}\caption{Summary of the behavior of three different approaches.}\label{tab:comp}
    \end{table}

\subsection{Inverse reconstruction of unknown fields and forward validation}\label{sec:invres}

\begin{figure}[h!]
    \centering
    \includegraphics[width=0.9\textwidth]{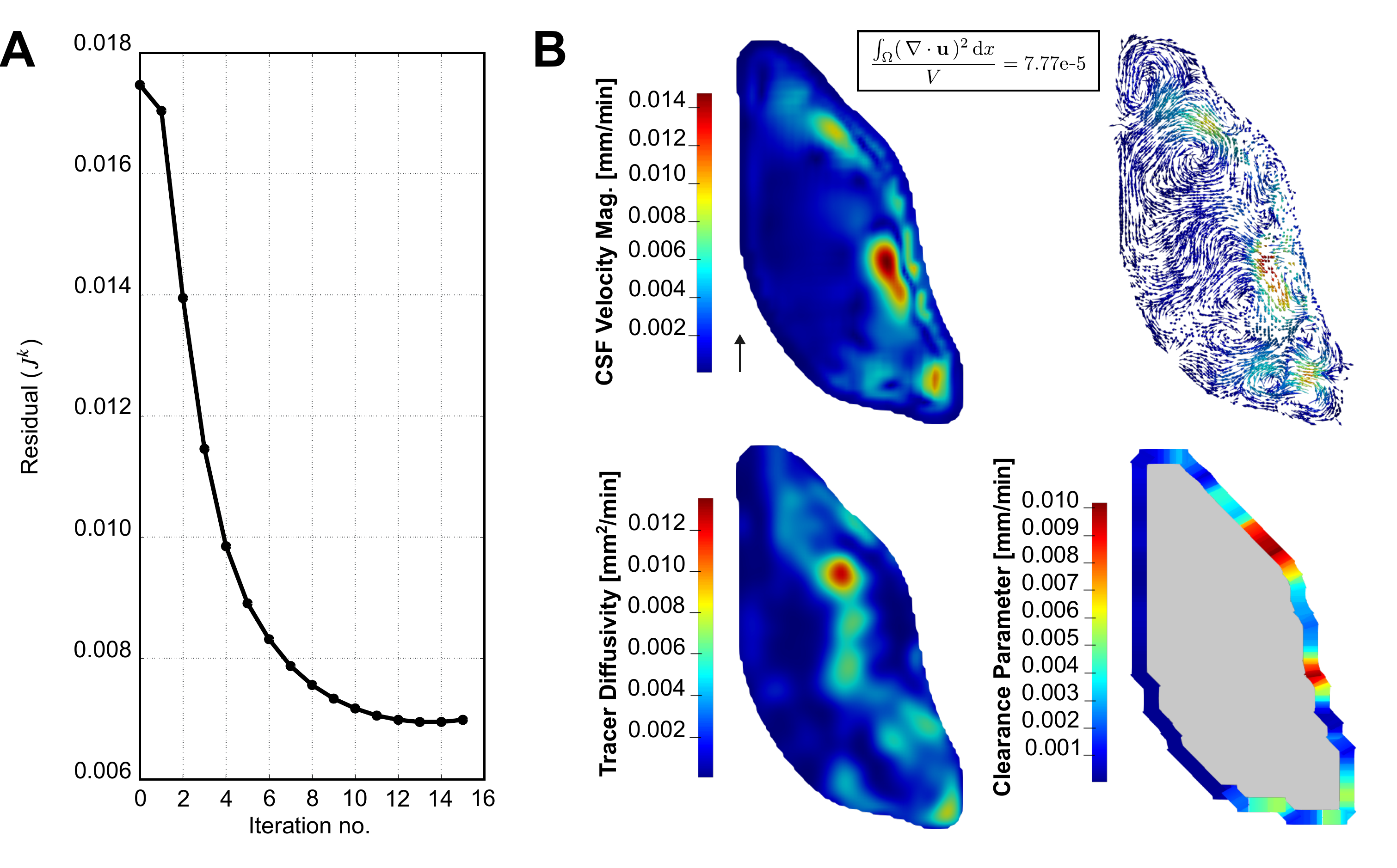}
    \caption{Panel A shows how the residual $J^k$ as defined in Eq.~(\ref{eq:min-J}) changes over iteration steps $\ell$ until a minimum is reached. Panel B displays the converged advective (top), diffusive (bottom) and clearance (bottom) fields. Arrow indicates anterior direction. The boxed quantity in Panel B carries the L$^2$ norm of the divergence of the converged advective field, normalized by the volume of the domain $V$.}
    \label{fig:iterations}
\end{figure}

Given the promise of the HW modified approach, we embed this formulation directly within an inverse problem framework to infer the velocity field $\ubar$.
This strategy replaces finite-difference estimation with a finite element-based inversion that better captures the underlying physics of the experimental observations.
For brevity, the implementation of the inverse problem with the alternative two methods is not included here. 
The advective, diffusive and clearance fields obtained by running the inverse problem minimization algorithm outlined in Sec.~\ref{sec:invform} are shown in Fig.~\ref{fig:iterations}.
Here, the calibration is done over the five timepoints shown in Fig.~\ref{fig:experiment}, and the initial conditions for the three fields have been selected based on literature values for the advective and clearance fields, and a scaling of the ADC map.
Once the residual remains unchanged over several successive iterations, as illustrated in Fig.~\ref{fig:iterations}A, the optimization is stopped.
Fig.~\ref{fig:iterations}B shows a mid-sagittal slice of the converged fields. 
The top row displays the obtained CSF velocity field, with the magnitudes (left) and vector map (right).
The inferred fields exhibit three key features consistent with glymphatic clearance. 
First,  compared with the velocity fields obtained in the previous section in Fig.~\ref{fig:ubar}, the velocity field here displays significant improvements, with peak velocities now appearing along the brain's inferior edges, where clearance is expected to be the highest~\cite{benveniste_glymphatic_2021}. 
Second, the vector field displays physically realistic, divergence-free flow patterns with vortices forming around major ventricles.
Third, the inferred diffusivity and clearance maps are consistent with the expected physics and assumptions related to the problem: (i) clearance is negligible along the superior skull boundary,
and (ii) clearance is significantly elevated near the base of the brain, which 
aligns with the findings reported in ~\cite{johnson_image-guided_2023}. 

\begin{figure}[h!]
    \centering
    \includegraphics[width=0.78\textwidth]{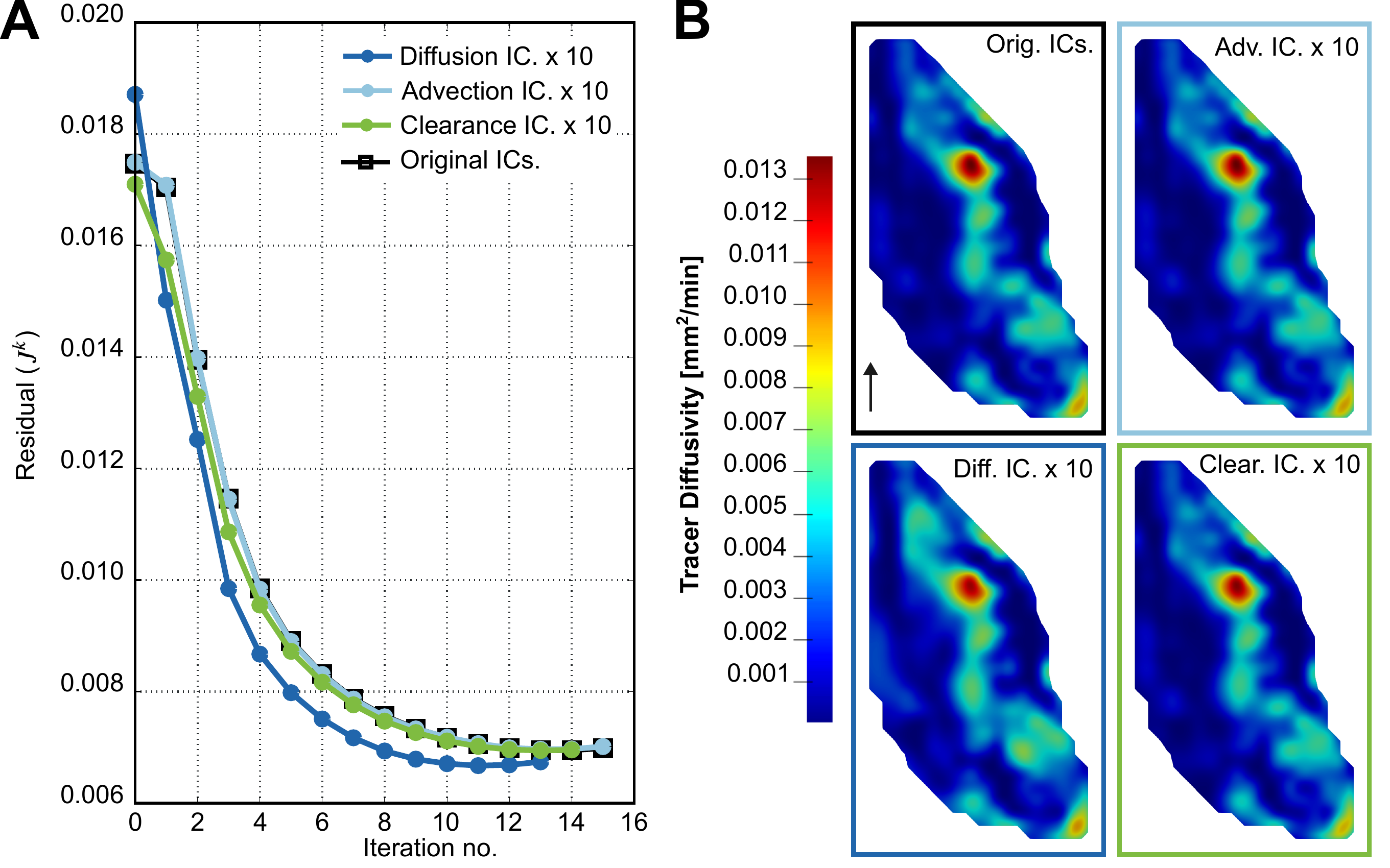}
    \caption{Panel A shows how the residual $J$ as defined in Eq.~(\ref{eq:min-J}) changes over iteration steps $\ell$ until a minimum is reached, for each run of the inverse problem with varying initial conditions (ICs). Panel B has the diffusivity field obtained with the four different initial conditions: with original ICs (top-left, same as Fig.~\ref{fig:iterations}), with advective ICs scaled by a factor of 10 (top-right), with diffusion ICs scaled by a factor of 10 (bottom-left) and with clearance ICs scaled by a factor of 10 (bottom-right). Arrow indicates anterior direction.}
    \label{fig:ICs}
\end{figure}

The sensitivity of the results in Fig.~\ref{fig:iterations} to the initial conditions are shown in Fig.~\ref{fig:ICs}, which displays the residual trajectory (inset A) and corresponding inferred fields (inset B) obtained with varying initial conditions (ICs). 
Each curve corresponds to a complete run of the inverse model in which the initial guess for the indicated field was scaled by a factor of 10, 
while the remaining parameters were held at their baseline values.
For example, the “Diffusion IC x 10” curve shows the residual plot for the optimization where the diffusion initial condition used in Fig.~\ref{fig:iterations} was multiplied by 10, and the advection and clearance initial conditions remained unchanged. 
The corresponding converged diffusion field, shown in the labeled inset of Fig.~\ref{fig:ICs}B, is representative of the results obtained across all tested initial conditions.
Notably, the converged fields remain nearly identical regardless of the starting point, exhibiting only negligible variations.
This stability is further evidenced by the residual trajectories in Fig.~\ref{fig:ICs}, which exhibit nearly complete overlap. 
Collectively, these observations suggest that the developed method is insensitive to perturbations in the initial conditions, provided they remain within the broad vicinity of the physically-informed values originally selected.

\subsubsection{Comparison with experimental CE-MRI data}

\begin{figure}[h!]
    \centering
    \includegraphics[width=0.9\textwidth]{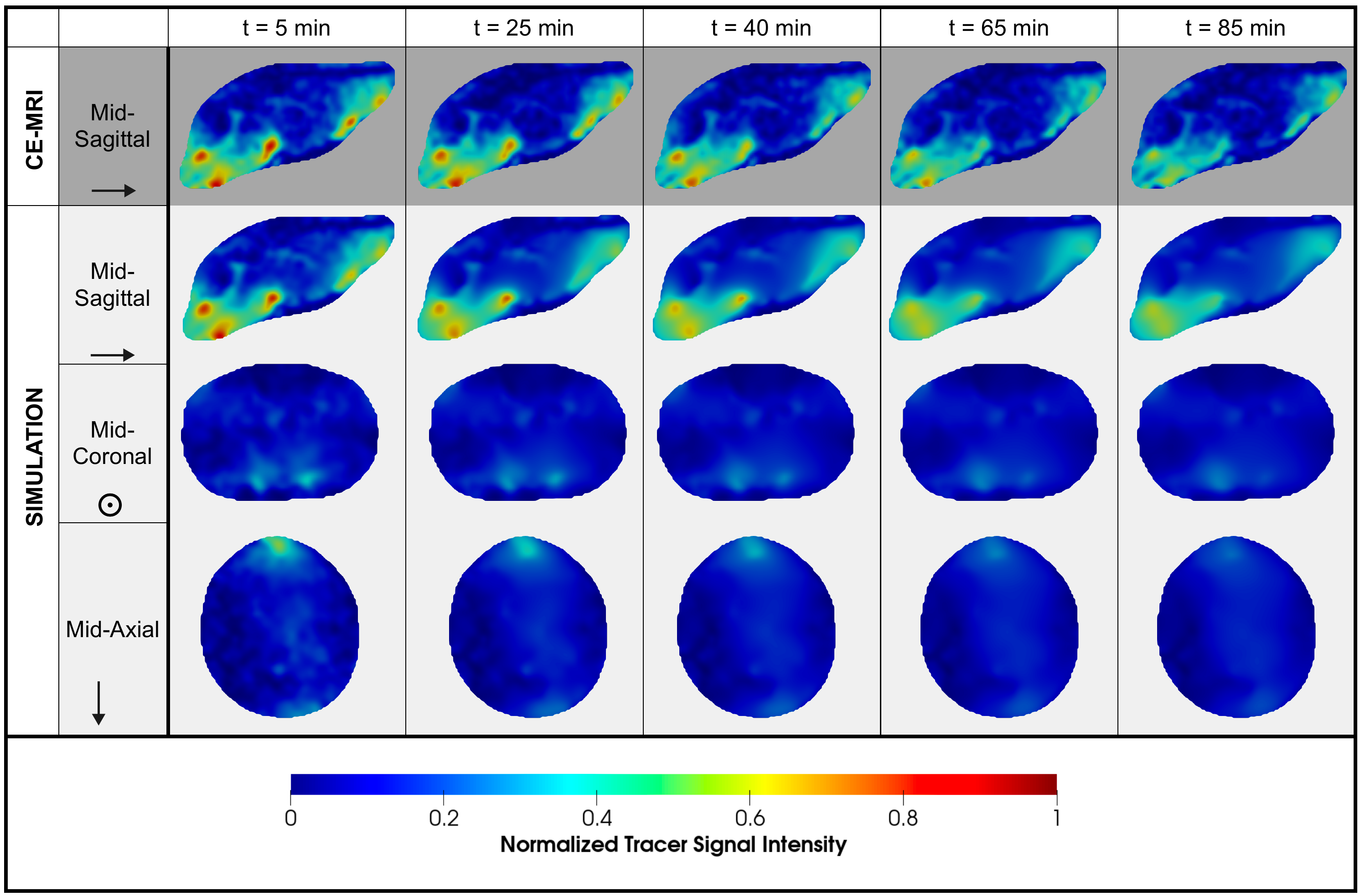}
    \caption{Qualitative comparison of the experimental data and forward simulation results obtained using converged fields from the HW modified-inverse approach. Top row corresponds to experimental CE-MRI images of the selected window, projected onto the B-spline mesh used for the numerical implementation. Bottom three rows display simulation results from three different slices of the brain. Here, arrows indicate the anterior direction.}
    \label{fig:table}
\end{figure}

Fig.~\ref{fig:table} compares the forward simulation results generated using the inferred fields from Fig.~\ref{fig:iterations} against the independent experimental validation set.
The initial condition at $t=0$ for the tracer distribution in the forward problem is the normalized tracer intensity from the first MRI acquisition time point, as illustrated within the region of interest in Fig.~\ref{fig:experiment}. 
Fig.~\ref{fig:table} displays the CE-MRI data projected onto the computational mesh in the top row, at the designated timepoints in each column, until $t=85$ minutes. 
The bottom three rows display the simulated tracer distributions at each timestep on mid-sagittal, mid-coronal, and mid-axial slices of the brain geometry. 
A qualitative comparison of the final timestep, specifically along the mid-sagittal plane, reveals strong agreement between the experimental observations and the simulation results.
Most importantly, the simulation captures the locations of both tracer accumulation and non-accumulation, as well as the relative intensity levels observed in those regions. 
 
\begin{figure}[h!]
    \centering
    \includegraphics[width=0.9\textwidth]{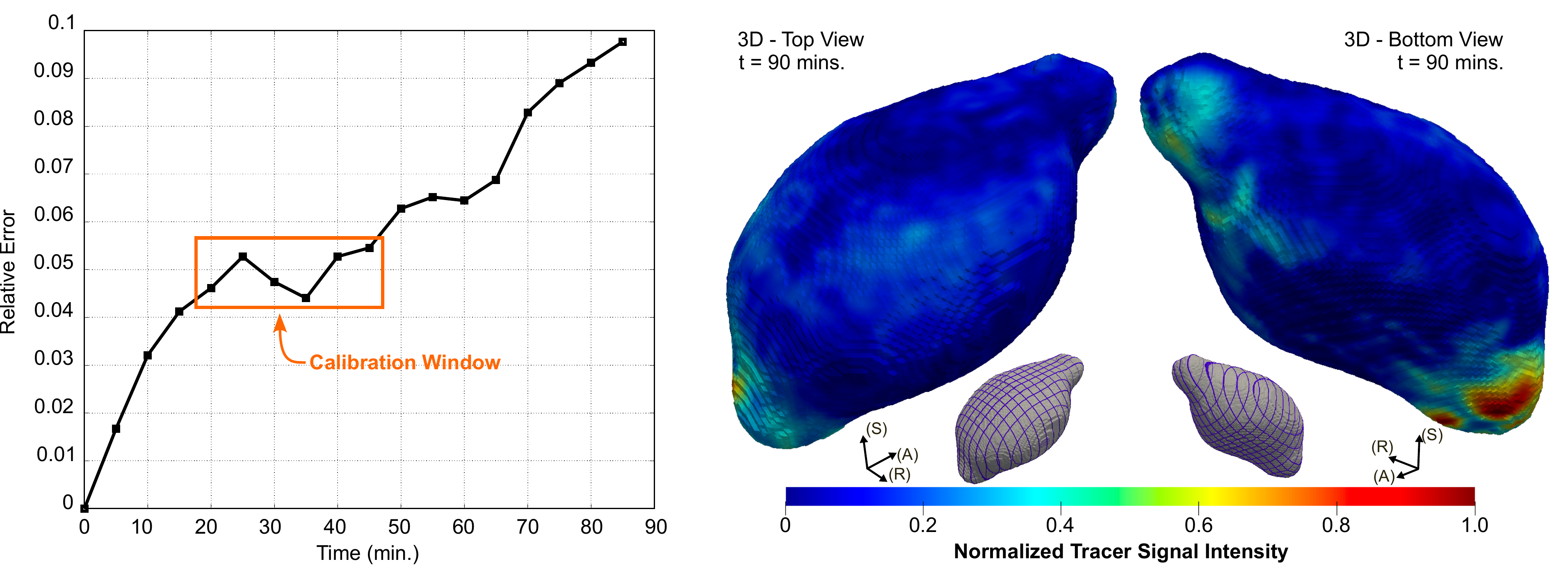}
    \caption{Quantitative comparison of experimental data and forward simulation results obtained using the converged fields from the HW modified-inverse approach (left). Relative error is calculated from Eq.~(\ref{eq:min-J}), normalized with the total image concentration at each step. 3D views of the concentration distribution at $t=90$ minutes are shown on the right. Coordinate directions (A: anterior, S: superior, R: right) are indicated.}
    \label{fig:Js}
\end{figure}

Finally, Fig.~\ref{fig:Js} provides a quantitative comparison of the experimental data and simulation results. 
The plot shows the relative error at each timestep $n$ calculated via~(\ref{eq:J-rel}), and highlights the region corresponding to the calibration data to demonstrate that the calibration data is a small subset of the whole set. 
The error increases as the simulation progresses, which is expected due to the diffusive nature of the problem, however; the highest error remains at $0.1$, which is much lower than any of the finite-difference results shown in Fig.~\ref{fig:compresults}. 
The right inset shows 3D views of the brain geometry and the concentration distribution at $t=90$ minutes, showing that the obtained results look physical all throughout the domain. 
The results combined demonstrate the utility of the approach outlined in this work and its ability to accurately capture the underlying physics of image data. 

Fig.~\ref{fig:Js} shows that the error in the simulation results approximating the experimental behavior increases with simulation time. 
This result is expected, because the calibration is performed over the indicated orange window of five time points. This means that the parameter fields are chosen to best approximate the result of a five-step forward run, compared to the experimental data at the last timestep of that window.

While this approach demonstrates the predictive capability of the framework, it remains to be determined whether calibrating the model against the full experimental dataset would yield superior results. This is of particular importance for future work incorporating amyloid plaque formation, a key pathological feature that will be essential for extending this inverse framework to neurodegenerative disease models.

\subsubsection{Sensitivity of inference on calibration window}

\begin{figure}[h!]
    \centering
    \includegraphics[width=0.9\textwidth]{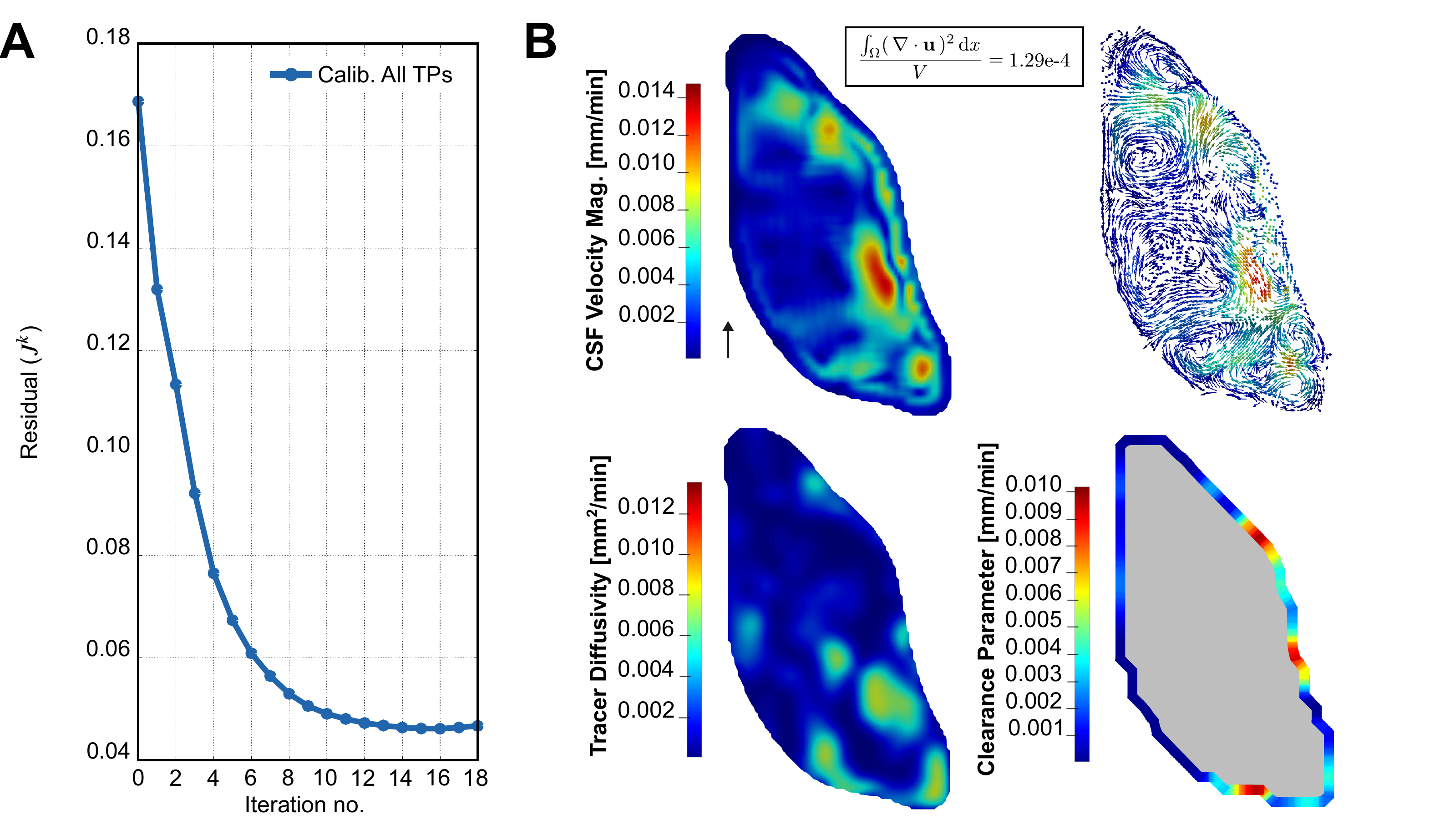}
    \caption{Panel A shows how the residual $J^k$ as defined in Eq.~(\ref{eq:min-J}) changes over iteration steps $\ell$ until a minimum is reached, \emph{using the whole set of glymphatic clearance data for the calibration}. Panel B displays the converged advective (top), diffusive (bottom left), and clearance (bottom right) fields obtained from this calibration. Arrow indicates anterior direction. The boxed quantity in Panel B carries the L$^2$ norm of the divergence of the converged advective field, normalized by the volume of the domain $V$.}
    \label{fig:iterations-all}
\end{figure}

Fig.~\ref{fig:iterations-all} shows the results of running the inverse algorithm using \emph{all the data points in the glymphatic clearance window} shown in Fig.~\ref{fig:experiment}, which constitute of 17 timesteps in total. 
Fig.~\ref{fig:iterations-all}A presents a higher floor for the values of the residual $J^k$ compared to those in Fig.~\ref{fig:iterations}.
This stems from the requirement to identify a consistent set of advection, diffusion, and clearance fields that can simultaneously match the experimental data across all time points, a task significantly more difficult than matching any five-point subset.
The converged fields in Fig.~\ref{fig:iterations-all} exhibit a spatial profile more closely aligned with physiological expectations 
compared to those in Fig.~\ref{fig:iterations}. 
Specifically, the advective field obtained using all the time points in the clearance window is more active around the base of the brain, where most clearance activity is assumed to take place.
Furthermore, the velocity maintains a divergence of similar order, ensuring the field remains weakly divergence-free.
We observe more distinct patterns in the clearance field, and a more subtle diffusion field. 
This indicates that as we widen the window for calibration, the influence of the diffusive field decreases while the advective field becomes more dominant in certain regions of the brain. 
Nevertheless, it should be noted that the transport remains diffusion-dominated across all calibrations, even as advective effects become more pronounced in certain regions.
This is further supported by solving the inverse problem for a family of models consisting of the full formulation in Eq.~(\ref{eq:weak-advec-HWC}), a diffusion-only model, and an advection-only model.
The diffusion-only model produced reconstructed fields and forward results comparable to the full model, except for its inability to capture sharper features at earlier timesteps.
In contrast, the inverse solver for the advection-only model failed to converge to any solution in the vicinity of those obtained for the other models, indicating that advection alone is insufficient to reproduce the observed transport.
All these findings together suggest that both advective and diffusive transport, acting at different scales, play significant roles in glymphatic clearance.

\begin{figure}[h!]
    \centering
    \includegraphics[width=0.9\textwidth]{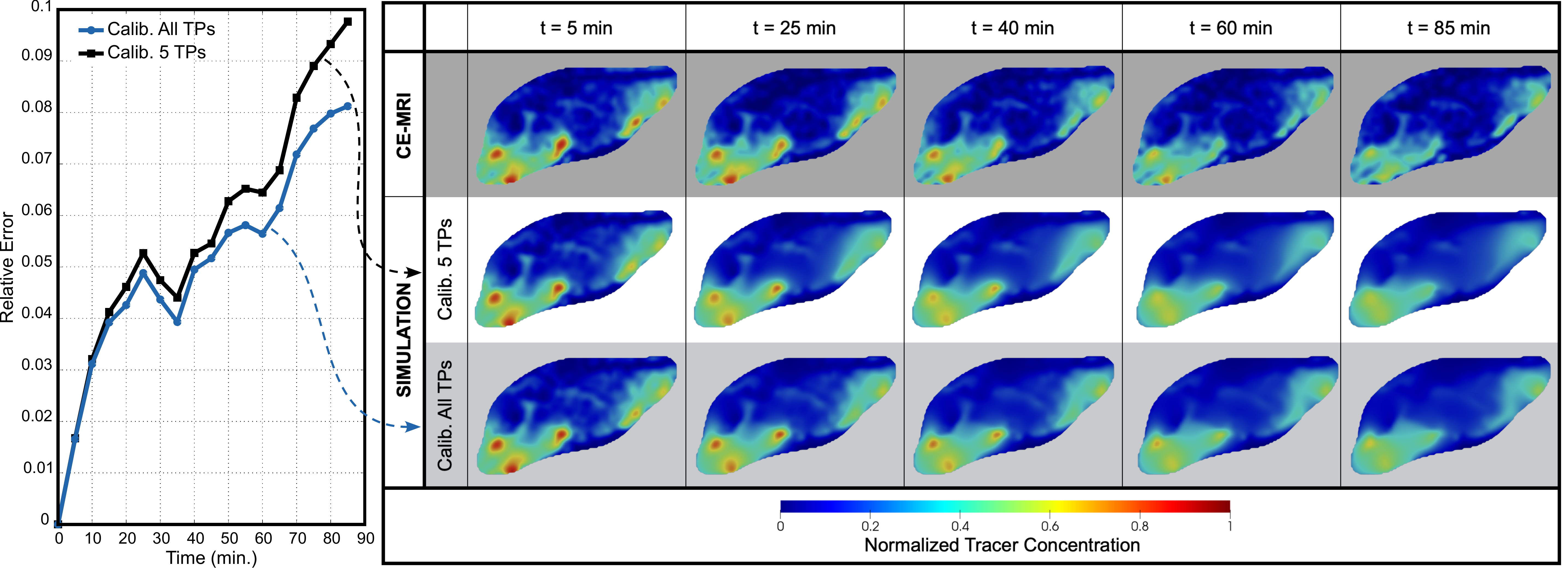}
    \caption{Quantitative and qualitative comparison of experimental data and forward simulation results obtained using the converged fields from the HW modified inverse approach, for two calibration strategies: using all the timepoint data for calibration, and using a subset of 5 timepoints for calibration. The plot shows the relative error over time, computed from Eq.~(\ref{eq:min-J}) and normalized with the total image concentration at each step, using either all time points or a reduced set of five calibration points. 
    Table compares the experimental CE-MRI data (top row) with the corresponding simulations obtained from the five-point calibration (middle row) and the full-time-point calibration (bottom row). 
    Each column corresponds to a different imaging time within the selected window. Tracer concentration is shown in normalized units}
    \label{fig:table-all}
\end{figure}

Fig.~\ref{fig:table-all} presents the forward-run for the previous calibration using five timepoints, together with the new calibration results using all timepoints, which converged to the fields in Fig.~\ref{fig:iterations-all}.
The graph plots the relative total error at each timestep between the simulated and experimental data.
Clearly, using the full range of data for calibration results in a better match between simulations and experimental data, especially at later time points, with the total error at the final time point being almost $20\%$ lower than before.
This improvement is also visible qualitatively in Fig.~\ref{fig:table-all};
the simulation that uses the calibrated data from all timepoints is significantly better at capturing certain peaks in the data, and remains noticeably less diffused as time progresses. 
The transport dynamics near the injection site are also captured more accurately, since the calibration now includes timepoints closer to the injection.
However, this is not necessarily desirable since the tracer injection represents a disturbance to the natural system, and it is the undisturbed system that we are trying to capture.

\begin{figure}[h!]
    \centering
    \includegraphics[width=0.7\textwidth]{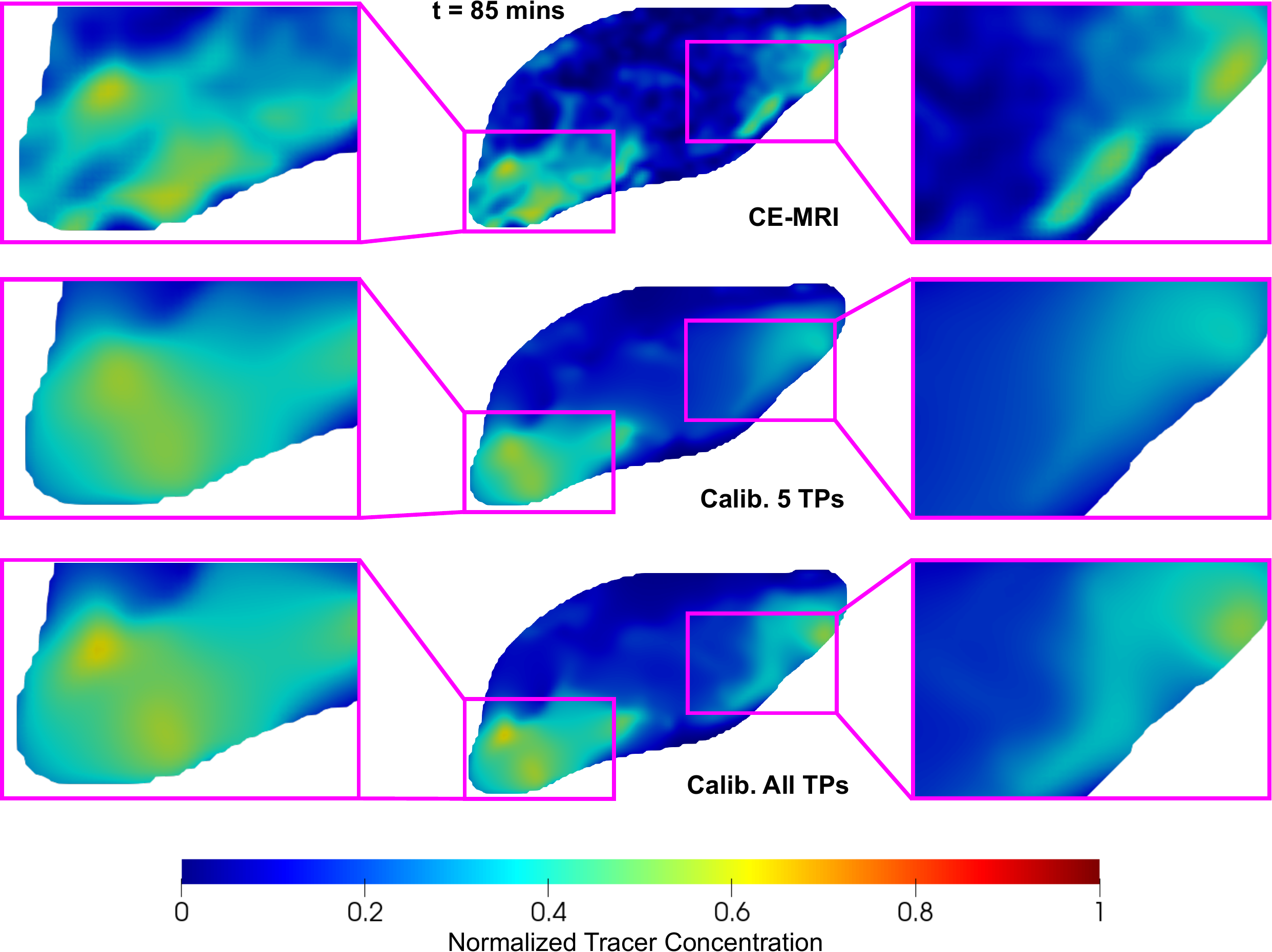}
    \caption{Close-up comparison of tracer concentration fields at the final time point ($t=85$ min). The experimental CE-MRI data (top) is compared with forward simulations obtained using the five-point calibration (middle) and the full-time-point calibration (bottom). Insets highlight selected regions to emphasize local differences between experimental and simulated fields for different calibration sets. Tracer concentration is shown in normalized units.}
    \label{fig:zoom-all}
\end{figure}

A closeup of the final timestep is shown in Fig.~\ref{fig:zoom-all}.
Comparing the two calibration strategies against the experimental data, we again observe that the simulation using all timepoints captures contours and patterns more accurately than the five-point calibration. 
Overall, using more timepoints has the advantage of (i) reducing the overall error across all timesteps, and (ii) producing less diffusive results better capturing sharp features.
However, using all time points in the clearance window requires solving a longer forward problem at each inverse iteration, which increases the computational cost.
Furthermore, for problems where the initial conditions are harder to determine or lack a clear physical interpretation, widening the calibration window may make convergence more difficult, since the choice of initial guess may not be equally suitable for both short and long windows.
Finally, it should be noted that the glymphatic flow considered here is a slow flow, where transport dynamics occur over hours rather than seconds or minutes, as in faster flows such as blood circulation. 
Thus, the assumption that the inferred fields are time-independent has a physical basis, and using more time points for calibration does not violate this assumption.
For flows that are expected to change significantly over the calibration window, however, widening the calibration window may not necessarily produce better results, and care must be taken in choosing a physically-motivated calibration interval. 

This proposed inverse modeling framework can easily be applied to the alternative two methods outlined in Sec.~\ref{sec:gen}, whose results are not presented here to avoid diverting the discussion. 
Using the inverse problem with the conservative or advective method yields results comparable to the HW modified advective inverse approach when appropriate measures are taken to ensure stability, however the HW modified advective inverse approach still renders the best overall match to experimental data.
Furthermore, in terms of the physicality and the biological interpretability of the results, the HW modified advective inverse approach is found to be far superior.

\subsubsection{Localized boundary efflux vs. global metrics of clearance}

Consensus on clearance definitions and values remains elusive in the current literature, characterized by notable discrepancies across experimental and computational studies. 
In experimental settings, a global clearance rate $k$  (1/min) is typically derived from the time evolution of total tracer mass in the domain. 
As an emergent, state-dependent quantity that treats the brain as a uniform sink, $k$ is inherently sensitive to the instantaneous spatial distribution and the initial injection site of tracer, which complicates direct comparisons between subjects or studies.
In glymphatic transport models, a volumetric reaction term ($rc$) is often added to the strong form~(\ref{eq:strong}) to account for clearance~\cite{vinje_human_2023,chen_unbalanced_2024}.
Here, $r$ (1/min) denotes a local reaction rate representing the absorption or appearance of the tracer; however, the specific biological mechanism this term is meant to represent remains physically ambiguous within the context of tracer experiments.
Physiologically, $r$ represents trans-vascular efflux where tracer is assumed to disappear into the local capillary bed via the blood-brain barrier (BBB) at every point in the tissue~\cite{smith2017}. 
Thus, in a comprehensive model of the transport of metabolic waste (e.g. amyloid), this term would account for local receptor-mediated absorption into the bloodstream. 
However, this volumetric representation does not align with the transport dynamics of our specific tracer experiment model.
In our mouse study, the tracer is administered via the cisterna magna, providing direct access to the cerebrospinal fluid pathways while largely bypassing the BBB~\cite{iliff_paravascular_2012}. 
We therefore assume volumetric effects are negligible in this context and instead characterize clearance as a predominantly surface phenomenon using a Robin-type boundary condition~(\ref{eq:BCs}),
thereby explicitly capturing the surface-dominated efflux occurring at the dural interfaces.

\begin{figure}[h!]
    \centering
    \includegraphics[width=0.9\textwidth]{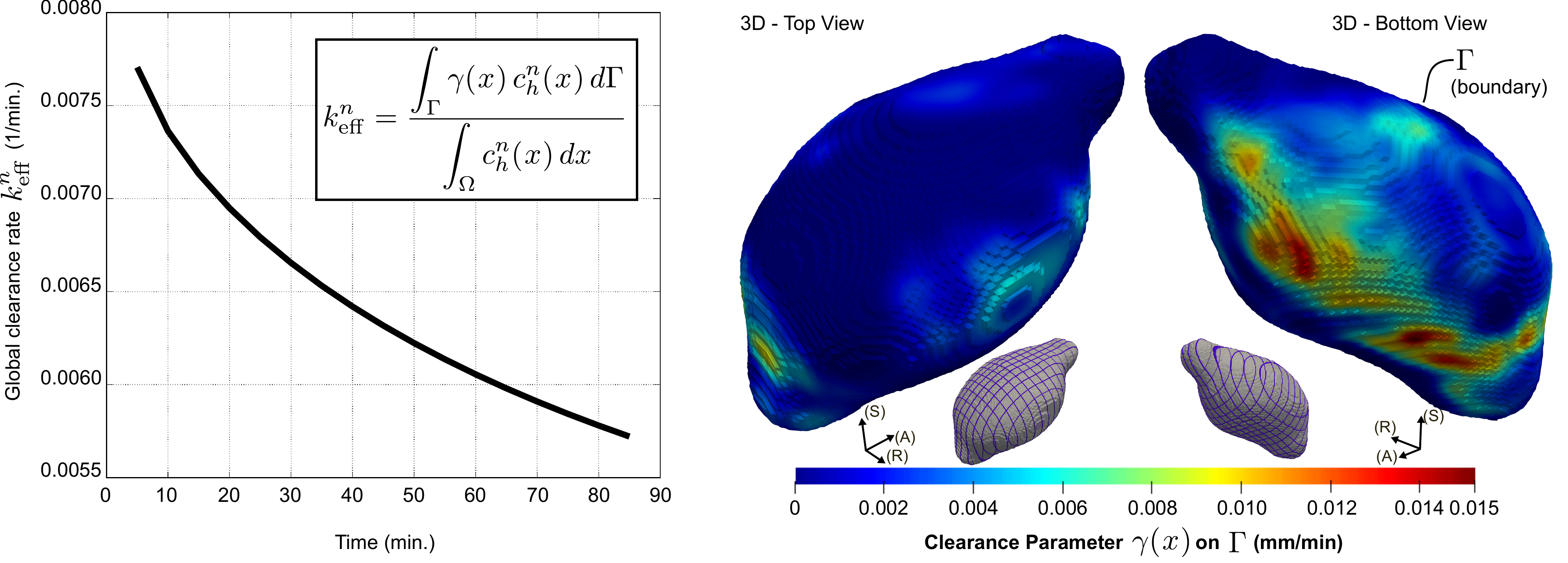}
    \caption{Comparison of emergent clearance rates and intrinsic boundary permeability. (Left) The effective global clearance rate $k_{\mathrm{eff}}^n$, calculated via Eq.~(\ref{eq:efflux}), demonstrates a clear time-dependence as the tracer’s spatial distribution evolves during the forward simulation. (Right) The inferred spatial distribution of the clearance parameter $\gamma(\boldsymbol{x})$ (mm/min) on the brain boundary $\Gamma$. Superior and inferior views reveal localized clearance ``hotspots'' in the brainstem and neck regions. Unlike the time-varying global rate, $\gamma$ represents an intrinsic, state-independent property of the drainage pathways. Coordinate directions (A: anterior, S: superior, R: right) are indicated.}
    \label{fig:efflux}
\end{figure}

The spatially varying clearance parameter $\gamma$ (mm/min) in our Robin-type boundary condition represents the \emph{local solute permeability} (or ``conductance'') of the brain’s outer membranes to the tracer. 
Physically, $\gamma$ provides a mechanistic measure of the solute efflux capacity, quantifying the ease with which metabolic waste can physically exit the parenchyma into the lymphatic pathways for further drainage~\cite{aspelund2015}.
Thus, the clearance parameter $\gamma$ characterizes an \emph{intrinsic physical property of the drainage anatomy}.
By inferring the spatially varying values of $\gamma$ from CE-MRI data, we can gain insight into locations of high and low drainage, distinguishing between largely "insulated" superior surfaces and "porous" inferior drainage hotspots.
This is clearly seen in our results for the spatially resolved $\gamma$ field, shown in Fig.~\ref{fig:efflux} (right), which suggests that high-clearance regions are concentrated in the inferior brain and neck.
This aligns with known anatomical exit routes, such as the cribriform plate and cranial nerves~\cite{ma2017}, and reflects the physical capacity of specific drainage pathways.

For completeness, and to facilitate comparison with existing works, we note that the present framework can also yield an equivalent global rate of clearance, $k_{\text{eff}}$ at any timestep $t^n$,
\begin{equation}\label{eq:efflux}
k_{\text{eff}}^n =
\f{ \int_{\Gamma} \gamma \, c^n_h(x)\, d\Gamma}{\int_{\Omega} c^n_h(x)\, dx}.
\end{equation}
In~(\ref{eq:efflux}), the numerator represents the total efflux rate (nM mm$^3$/min) accross the boundary $\Gamma$ (total amount of tracer leaving the domain per unit time) at timestep $t^n$, while the denominator represents the total integrated tracer mass within the domain ${\Omega}$ at timestep $t^n$, with units of nM mm$^3$.
Their ratio defines $k_{\text{eff}}^n$ as a scalar rate with units of $1/$min.
Fig.~\ref{fig:efflux} illustrates this global clearance rate alongside the inferred 3D map of the clearance parameter $\gamma$.
While the derived quantity $k_{\text{eff}}$ is a transient metric that evolves based on the tracer distribution, the spatial map of $\gamma$ provides a time-invariant, image-inferred signature of the subject's clearance architecture. Unlike global rates that are sensitive to the instantaneous concentration field, and thus the specific tracer injection site, $\gamma$ represents a mechanistic quantity independent of the tracer’s spatiotemporal state, offering a more robust basis for physiological interpretation and cross-subject comparison. Note, while $\gamma$ is treated as a time-invariant parameter in this study consistent with the stable anatomical state of the brain during a 90-minute imaging session, the framework is mathematically extensible to time-varying fields. Such an extension would be pertinent for modeling transient physiological events, such as the circadian fluctuations known to influence glymphatic flux~\cite{Xie2013}.

\section{Conclusion}\label{sec:conc}

We have presented a modified formulation of the advection–diffusion equation that preserves conservation when used with experimentally obtained non-solenoidal advective fields.
The framework is based on an additive decomposition of the velocity, which constructs a corrective component for a given, non-physical velocity field obtained from experimental data or auxiliary simulations.
This forward formulation is advantageous in applications where the provided velocity field is prescribed directly and may not satisfy the divergence-free condition 
inherent in incompressible flow.
The accompanying inverse problem framework allows the unknown, physically grounded transport fields to be recovered from experimental data using Newton iterations.
Within this framework, not only the advective field but also other transport parameter fields, such as diffusion and clearance, can be simultaneously inferred.
The additive decomposition, together with an IGA  implementation on B-spline bases, provides a natural mechanism for field regularization and for mitigating the effects of image noise.
Consequently, the recovered velocity fields weakly satisfy the divergence-free condition, yielding stable and physically consistent transport predictions.
While this study is focused on glymphatic clearance and its potential links to neurodegenerative diseases such as Alzheimer’s disease, the methodology remains broadly applicable to biological systems where incompressible flow must be inferred from experimental imaging.
Future work will focus on refining the selection of calibration data, exploring the effects of time-varying parameter fields, and incorporating amyloid plaque formation into the reconstruction to predict long-term deposition patterns, which is accommodated by the model we have presented in~\cite{johnson_image-guided_2023}.

\section*{Acknowledgments}
\addcontentsline{toc}{section}{Acknowledgments}
Research reported in this publication was supported by the National Institute on Aging (NIA) of the National Institutes of Health (NIH) under Award Number R21AG080207.

\clearpage\newpage

\appendix

\section{Derivation of inverse operators}\label{sec:app}

The details of the terms involved in solving the inverse problem are given here. 
We can obtain an expression for $\f{\p L_{ij}}{\p \, \b{\Theta}} \, \alpha^{\, n+1}_j$ by defining the derivatives of each operator in $L_{ij}$ with respect to the parameters in $\b{\Theta}$,
\begin{equation}\label{eq:dLijdT}
    \begin{aligned}
       \f{\p L_{ij}}{\p \, \Theta_m}\, \alpha^{\, n+1}_j &= \Delta t \, \Bigg[\, \f{\p K_{ij}}{\p \, \Theta_m} + \f{\p \Abar_{ij}}{\p \, \Theta_m} + \f{\p A^{\prime}_{ij}}{\p \, \Theta_m} + \f{\p H_{ij}}{\p \, \Theta_m}\, \Bigg]\, \alpha^{\, n+1}_j\,. \\
    \end{aligned}
\end{equation}
Since the advection, diffusion, and boundary clearance operators act independently, each operator has nonzero derivatives only with respect to its own parameter set in $\boldsymbol{\Theta}$.
This allows us to reconstruct Eq.~(\ref{eq:dLijdT}) as
\begin{equation}\label{eq:dLijdT-contract}
    \begin{aligned}
       \f{\p L_{ij}}{\p \, \Theta_m}\, \alpha^{\, n+1}_j &= \Delta t \, \Bigg\{\, \f{\p K_{ij}}{\p \, D^{\,h}_m}\, \alpha^{\, n+1}_j \, , \, \f{\p \Abar_{ij}}{\p \,u^{\,h}_m}\, \alpha^{\, n+1}_j \, , \, \f{\p A^{\prime}_{ij}}{\p \, u^{\,h}_m}\, \alpha^{\, n+1}_j \, , \, \f{\p H_{ij}}{\p \, \gamma^{\,h}_m}\, \alpha^{\, n+1}_j\, \Bigg\}^{\,t} \,. \\
    \end{aligned}
\end{equation}
Using the definitions of the operators given in Eq.~(\ref{eq:Wh-ops}), together with the discretization definitions of the unknown fields~(\ref{eq:disc-inv-params}), we obtain expressions for the derivatives of each operator. 
The derivative of the diffusion operator $K_{ij}$ with respect to the unknowns in the diffusivity field $D^{\,h}$ reads
\begin{equation}\label{eq:KijDeriv}
    \begin{aligned}
       \f{\p K_{ij}}{\p \, \b{\Theta}}  \equiv \f{\p K_{ij}}{\p \, D^{\,h} \,} = \f{\p \, \Big(\,\intom D \, \nabla \, N_i \cdot \nabla N_j \, \d x \,\Big)}{\p \, D^{\,h} \,} \,   
        &=\f{\p \,\Big(\,\intom D^{\,h}_k N_k \, \nabla \, N_i \cdot \nabla N_j \, \d x \,\Big)}{\p \, D^{\,h}_m \,} \b{e}_i \dyad \b{e}_j \dyad \b{e}_m \, \\
        &= \intom \delta_{km} \, N_k \, \nabla \, N_i \cdot \nabla N_j \, \d x \, \b{e}_i \dyad \b{e}_j \dyad \b{e}_m \, \\
        &= \intom \, \nabla \, N_i \cdot \nabla N_j \, N_m \, \d x \, \b{e}_i \dyad \b{e}_j \dyad \b{e}_m \,.\\        
    \end{aligned}
\end{equation}
The derivative of the advection operator $\Abar_{ij}$ with respect to the advective field unknowns $\b{u}^{\,h}$ reads
\begin{equation}\label{eq:AbarijDeriv}
    \begin{aligned}
        \f{\p \Abar_{ij}}{\p \, \b{\Theta}} \equiv \f{\Abar_{ij}}{\p \, \b{u^{\,h}}} = \f{\p \,\Big(\,\intom N_i \, (\ubar \cdot \nabla N_j\,) \, \d x \,\Big)}{\p \, \b{u^{\,h}}} &= \f{\p \,\Big(\,\intom N_i \, N_{j,k}\, \ubar_k  \, \d x \Big)}{\p \, u^{\,h}_m} \,\b{e}_i \dyad \b{e}_j \dyad \b{e}_m \\
        &= \f{\p \,\Big(\,\intom N_i \, N_{j,k}\, N^{\,*}_{nk} u^{\,h}_n  \, \d x \Big)}{\p \, u^{\,h}_m} \,\b{e}_i \dyad \b{e}_j \dyad \b{e}_m \\
        &= \,\intom \delta_{mn} \, N_i \, N_{j,k}\, N^{\,*}_{nk}  \, \d x \,\b{e}_i \dyad \b{e}_j \dyad \b{e}_m \\
        &= \,\intom \, N_i \, N_{j,k}\, N^{\,*}_{mk}  \, \d x \,\b{e}_i \dyad \b{e}_j \dyad \b{e}_m \,.
    \end{aligned}
\end{equation}
To find the derivative of the second advection operator, we must first define $\Apri$ in terms of $\ubar$.
Recalling the definition of $\Apri$, and using Eq.~(\ref{eq:mass-cons}) provides a new way of writing $\Apri$ 
\begin{equation}\label{eq:Apri-ubar-2}
    \begin{aligned}
        \Apri_{ij} = \intom N_i \, N_j \, (\nabla \cdot \ubar) \, \d x \,,
    \end{aligned}
\end{equation}
whose derivative with respect to $\b{\Theta}$ we can now calculate as
\begin{equation}\label{eq:ApriijDeriv}
    \begin{aligned}
        \f{\p \, \Apri_{ij}}{\p \, \b{\Theta}} \equiv \f{\p \, \Apri_{ij}}{\p \, \b{u^{\,h}}} = \f{\p \,\Big(\,\intom N_i \, N_j \nabla \cdot \ubar \, \d x \,\Big)}{\p \, \b{u^{\,h}}} &= \f{\p \,\Big(\,\intom N_i \, N_{j}\, \ubar_{k,k}  \, \d x \Big)}{\p \, u^{\,h}_m} \,\b{e}_i \dyad \b{e}_j \dyad \b{e}_m \\
        &= \f{\p \,\Big(\,\intom N_i \, N_{j}\, N^{\,*}_{nk,k} u^{\,h}_n  \, \d x \Big)}{\p \, u^{\,h}_m} \,\b{e}_i \dyad \b{e}_j \dyad \b{e}_m \\
        &= \,\intom \delta_{mn} \, N_i \, N_{j}\, N^{\,*}_{nk,k}  \, \d x \,\b{e}_i \dyad \b{e}_j \dyad \b{e}_m \\
        &= \,\intom \, N_i \, N_{j}\, N^{\,*}_{mk,k}  \, \d x \,\b{e}_i \dyad \b{e}_j \dyad \b{e}_m \,.
    \end{aligned}
\end{equation}
Finally, the derivative of the clearance operator reads
\begin{equation}\label{eq:HijDeriv}
    \begin{aligned}
       \f{\p H_{ij}}{\p \, \b{\Theta}}  \equiv \f{\p H_{ij}}{\p \, \gamma^{\,h}} = \f{\p \, \Big(\,\int_{\Gamma} \, \gamma \, N_i \, N_j \, \d x \,\Big)}{\p \,\gamma^{\,h} \,} \,   
        &=\f{\p \,\Big(\,\int_{\Gamma} (\gamma^{\,h}_k N_k) \, N_i \, N_j \, \d x \,\Big)}{\p \, \gamma^{\,h}_m \,} \b{e}_i \dyad \b{e}_j \dyad \b{e}_m \, \\
        &= \int_{\Gamma} \delta_{km}  N_k  \, N_i \, N_j \, \d x \, \b{e}_i \dyad \b{e}_j \dyad \b{e}_m \, \\
        &= \int_{\Gamma}  N_i \, N_j \, N_m \, \d x  \, \b{e}_i \dyad \b{e}_j \dyad \b{e}_m \,. \\        
    \end{aligned}
\end{equation}
Collecting the obtained expressions for the derivatives of all operators,
\begin{equation}\label{eq:dKdTs}
    \begin{aligned}
        \f{\p \, K_{ij}}{\p \, D^{\,h}_m} \, \alpha_j^{n+1} &= \intom \, N_{i,k} \, N_m \, N_{j,k} \,\alpha^{n+1}_j \, \d x \, \b{e}_i \dyad \b{e}_m =: k_{im} \,, \\
        \f{\p \, \Abar_{ij}}{\p \, u^{\,h}_m} \, \alpha_j^{n+1} &= \intom \, N_i \, N^{\,*}_{mk} \, N_{j,k} \, \alpha_j^{n+1}   \, \d x \,\b{e}_i \dyad \b{e}_m =: \overline{a}_{im} \,, \\
        \f{\p \, \Apri_{ij}}{\p \, u^{\,h}_m} \, \alpha_j^{n+1} &= \intom \, N_i \, N^{\,*}_{mk,k}  \, N_{j} \, \alpha_j^{n+1}  \, \d x \,\b{e}_i \dyad \b{e}_m =: a^\prime_{im} \,, \\
        \f{\p \, H_{ij}}{\p \, \gamma^{\,h}_m} \, \alpha_j^{n+1} &= \int_{\Gamma} \, N_i \, N_m \, N_{j} \, \alpha_j^{n+1} \d \Gamma \, \b{e}_i \dyad \b{e}_m =: h_{im} \,.
    \end{aligned}
\end{equation}
We can now rewrite Eq.~(\ref{eq:Wh-simp-diff}) in terms of the new operators~(\ref{eq:dKdTs}) as
\begin{equation}\label{eq:Wh-simp-diff-simp}
    \begin{aligned}
        & L_{ij} \, \f{\p \, \alpha_j^{n+1}}{\p \Theta_m}  = \Big[ \, M_{ij} + M^{\,s}_{ij}\, \Big] \, \f{\p \alpha_j^{n}}{\p \Theta_m} \, - \Delta t \, \Bigg{\{}\, k_{im} \, , \, ( \, \overline{a}_{im} + a^\prime_{im} \, ) \, , \, h_{im} \, \Bigg{\}}^t \,,
    \end{aligned}       
\end{equation}
which renders a system where we can solve for $\f{\p \, \alpha_j^{n+1}}{\p \Theta_m}$ at each timestep $n$, given $\f{\p \alpha_j^{n}}{\p \Theta_m}$ from the previous timestep, and the known operators $k_{im}$, $\overline{a}_{im}$, $a^\prime_{im}$, $h_{im}$, which include the solution of the forward problem at the current timestep.

\bibliographystyle{elsarticle-num}

\clearpage\newpage

\singlespacing
\bibliography{Glymphatics}

\end{document}